\documentclass[%
 reprint,
superscriptaddress,
 amsmath,amssymb,
 aps,
 longbibliography
]{revtex4-1}

\usepackage{mathtools}

\DeclarePairedDelimiter\floor{\lfloor}{\rfloor}

\usepackage[usenames]{color}
\usepackage{amssymb}
\usepackage{amsmath} 

\usepackage{esdiff}
\usepackage{amsthm}
\usepackage{placeins}
\usepackage{makeidx}
\usepackage{url}
\usepackage{float}
\usepackage[caption = false]{subfig}
\usepackage[normalem]{ulem}
\usepackage{tabto}
\usepackage{xcolor}
\usepackage{mathtools}

\usepackage{graphicx}
\usepackage{dcolumn}
\usepackage{bm}

\usepackage{etoolbox}
\makeatletter
\patchcmd\linenumberpar{\@LN@parpgbrk}{\penalty\@LN@parpgpen\relax}{}{}
\makeatother

\def\be{\begin{equation}}
\def\ee{\end{equation}}
\def\bea{\begin{eqnarray}}
\def\eea{\end{eqnarray}}

\newcommand{\ie}{\textit{i.e. }}
\newcommand{\eg}{\textit{e.g. }}

\definecolor{britishracinggreen}{rgb}{0.0, 0.26, 0.15}

\newcommand{\cg}[1]{#1}

\newcommand{\piero}[1]{#1}

\begin{document}

\title{
  \cg{Non-Markovian temporal networks with auto- and cross-correlated link dynamics}
}

\author{Oliver E. Williams}
\email{o.edgar.williams@gmail.com}
\affiliation{School of Mathematical Sciences, Queen Mary University of London, London, E1 4NS, United Kingdom}

\author{\cg{Piero Mazzarisi}}
\email{piero.mazzarisi@sns.it}
\affiliation{Scuola Normale Superiore, Piazza dei Cavalieri, 7, 56126 Pisa, Italy}

\author{Fabrizio Lillo}
\email{fabrizio.lillo@unibo.it}
\affiliation{Department of Mathematics, University of Bologna, Piazza di Porta San Donato 5, 40126, Bologna, Italy}
\affiliation{Scuola Normale Superiore, Piazza dei Cavalieri, 7, 56126 Pisa, Italy}

\author{Vito Latora}
\email{v.latora@qmul.ac.uk}
\affiliation{School of Mathematical Sciences, Queen Mary University of London, London, E1 4NS, United Kingdom}
	\affiliation{Dipartimento di Fisica ed Astronomia, Universit\`a di Catania and INFN, I-95123 Catania, Italy}
	\affiliation{Complexity Science Hub Vienna (CSHV), A-1080 Vienna, Austria}

\date{\today}

\begin{abstract}

Many of the biological, social and man-made networks around us are inherently dynamic, with their links switching on and off over time. The evolution of these networks is often non-Markovian, and the dynamics of their links correlated. Hence, to accurately model these networks, predict their evolution, and understand how information and other quantities propagate over them, the inclusion of both memory and dynamical dependencies between links is key. We here introduce a general class of models of temporal networks based on discrete autoregressive processes. As a case study we concentrate on a specific model within this class, generating temporal networks with a specified underlying backbone, and with precise control over the dynamical dependencies between links and the strength and length of their memories. In this network model the presence of each link is influenced by its own past activity and the past activities of other links, as specified by a coupling matrix, which directly controls the causal relations and correlations among links. We propose a method for estimating the models parameters and how to deal with heterogeneity and time-varying patterns, showing how the model allows for a more realistic description of real world temporal networks and also to predict their evolution. We then investigate the role that memory and correlations in link dynamics have on processes occurring over a temporal network by studying the speed of a spreading process, as measured by the time it takes for diffusion to reach equilibrium. Through both numerical simulations and analytical results, we are able to separate the roles of autocorrelations and neighbourhood correlations in link dynamics, showing that the speed of diffusion is non-monotonically dependent on the memory length, and that correlations among neighbouring links can speed up the spreading process, while autocorrelations slow it down.

\end{abstract}

\maketitle

\section{Introduction}

Much of the world we experience is governed by interactions. Networks
provide a natural way of modelling these interactions, and as such the
study of networks has been central to the understanding of both
natural phenomena and man-made systems.  Observably, many of the
networks around us change over time, as the interactions and
connections that define them come and go.  Human contacts and social
interactions do not last forever
\cite{Gonzalez08,Starnini:2013,Yoneki:2009}, roads between towns and
cities can be closed or new ones build
\cite{murcio2015multifractal,li2015percolation}, financial or economic
agents trade each day with different counterparts
\cite{mazzarisi2019dynamic}, and even our brains undergo significant
changes throughout our lives
\cite{Valencia08,Fallani08,millan2018concurrence,chialvo2010emergent}.
Real-world examples of temporal networks are often found to have a set
of very well defined structural and temporal features, many of which
play key roles in determining the dynamics and functioning of the
systems for which they form the backbone
\cite{Grindrod_prslA10,Holme_rev12,masuda_guide_temp_net,gauvin2014detecting,zanin2009dynamics,nicosia2012components,weng2017memory,peixoto2017modelling}.
Various models have been recently proposed to replicate such
features. For instance, models of human face-to face interactions
often rely on the assumption that the agents move as random walkers in
a physical space and create a link whenever they are closer than a
certain distance \cite{Starnini:2013,buscarino2008disease}.  Other
models take a slightly more abstract approach, introducing the notion
of node activity to control the presence of links
\cite{starnini2014temporal,karsai2014time,alessandretti2017random}. The
adaptations and extensions of these models do directly specify the
presence of empirically observed features such as memory, by which we
here mean a dependence on some finite number of past states. Indeed,
memory has been seen to play an important role in many real-world
networks
\cite{Singer_PLOSONE14,Fallani08,Szell:2012aa,lambiotte2019networks}.
It can affect the dynamics of social interactions
\cite{moinet2018generalized} and the controllability of temporal
networks \cite{zhang2017controllability}, and can also turn useful in
the definition of flow based communities
\cite{salnikov2016using,matamalas2016assessing,Rosvall_natcomm14}.  An
area of study in which memory has received a large attention is its
relation to spreading processes
\cite{Hiraoka:2018aa,sapienza2018estimating}. When considering the
spreading of an infection over a network, the presence of memory in
the link activities can have a considerable effect on the rate of
spreading of the disease, and can even cause dramatic changes to the
epidemic threshold
\cite{kiss2015generalization,peixoto2018change,Williams_2019,moinet2018effect}.
In diffusive procesess, memory directly induces the slow-down, or
speed-up, of the spread of information over the network
\cite{masuda2013temporal,delvenne2015diffusion,Hiraoka:2018aa,Rosvall_natcomm14,Scholtes_natcomm14,zhan2019information}.
This has been studied in the context of higher-order networks, and is
often understood to be a result of the correlated bursts, and the
induced lasting interactions that the non-exponential inter-event
times, which define memory, necessitate
\cite{jo2015correlated,Hiraoka:2018aa,burioni2017asymptotic,kim2015scaling,georgiou2015solvable,scholtes2017network}.
What has been done, however, does not form a full picture.
The presence of memory in the links that make up a network naturally
means that the state of each of these links at a given 
time can depend on the past activity of the link. It is common in
real networks to have pairs of links which are correlated
with each other. Indeed, it seems natural to assume that links in a 
temporal network can have memory of each others past, rather than
simply their own.  The connections between the rate at which
information spreads across a network and the memory of links are deep,
as are the connections between memory and link correlations. However,
the way in which inter link correlations and memory interact, and the
effects this interaction has on spreading and other dynamical
processes occurring over temporal networks is not well understood.

\cg{The goal of this article is twofold. We first introduce \piero{a novel and general class of models of temporal networks, which are based on a discrete autoregressive mechanism for link dynamics. Then, as a case study,  
we concentrate on a specific} 
  generative model for temporal networks \piero{within this class} in which the backbone
  structure, temporal correlations and memory are all taken into
  account, but can be precisely and separately controlled. We will also
  present a method for inferring the key parameters of the model from
  empirical data, simultaneously highlighting both time the ability of the
  model to describe real systems, and the role played by both memory
  and cross-interactions of links in the dynamics of real-world
  networks and in the forecasting of links. \piero{Then, we extend further the range of applicability by showing how to account for heterogeneity and time-varying patterns in link dynamics, again validating the proposed generalization on data.}
  The second goal of the
  article is to investigate how the interplay of the three key
  properties of a temporal network, namely the structure of its
  underlying backbone, the correlations between the evolution of its
  links, and the memory of its own past states, impact dynamical
  processes over the network.  In particular, we will study the way in
  which these properties affect a process of diffusion over a temporal
  network.}

This paper is organised as follows. \piero{In Section \ref{darns} we
  introduce a general }
  \piero{class of
  models of temporal networks based on discrete autoregressive
  processes. As a concrete case,} in Section \ref{section2}
  \piero{we consider a specific model
  within this class that allows a controlled description and treatment
  of the cross-interactions in link dynamics},
\cg{the so-called {\em Correlated Discrete Auto-Regressive Network}
  model of order $p$, or in short the CDARN($p$) model. We
  discuss our model in the context of other existing generative models
  for temporal networks, and we explain how the controllability,
  flexibility and analytical tractability of the model fills an
  important gap in the literature.
In Section \ref{empcdarn}, we show how the CDARN($p$) can be applied
to model real temporal networks presenting a maximum likelihood
estimation framework to infer the key parameters of the model from
empirical data. In this section the role played by both memory and
cross-interactions of links in the dynamics of real-world networks
will be evident, as well as the ability of the CDARN($p$) model to
effectively reproduce real features of temporal networks. Hence, we point out
that including cross interactions allows us to better describe the
evolution of real-world temporal networks, specifically by better predicting the
appearance of a link between a given couple of nodes. \piero{Moreover, we show how heterogeneous or time-varying parameters can be considered in our setting thanks to the flexibility of maximum likelihood approach. In particular we show the role played by both heterogeneous and time-varying patterns in estimation of and forecasting with the CDARN(p) model.}
In Section \ref{diff_process}, we consider processes occurring over temporal
networks. As an example of a network process, we study diffusion over
temporal networks generated by the the CDARN($p$) model.  We implement
the CDARN($p$) model on a number of backbone topologies taken from
real-world systems, and we present numerical and analytical results
concerning how the various features of the temporal network affect the
diffusion process occurring over it. In particular, we show} that the
average time taken for diffusion to reach equilibrium on these
networks is generally non-monotonically dependent on the memory
length, in agreement with recent findings regarding a different type
of process, namely epidemic spreading in temporal networks with only
self-correlated links activities \cite{Williams_2019}.  Here, however,
we find that the time taken to reach equilibrium is additionally
highly dependent on how links in the temporal network are
correlated. \cg{Moreover, and more importantly, we study in detail the
effects of link cross correlations on diffusion. We are able to
explain the dependence of the time to reach equilibrium on the types of
correlations between the activities of links in the temporal network.
Specifically we find that when correlations between neighbouring links
are strengthened when compared to link autocorrelations, diffusion
speeds up.}  This is a surprising complement to some recent works:
while autocorrelation in links slows down diffusion, as explained by
the induced burstiness of the link processes, correlations between
neighbour links speeds it up
\cite{delvenne2015diffusion,Hiraoka:2018aa,jo2015correlated,Lambiotte_jcn15,masuda2013temporal,Rosvall_natcomm14,Scholtes_natcomm14,Vestergaard_2014,Williams_2019,PhysRevE.92.012817}.

Overall, our results demonstrate that the topology of a temporal
network interacts in a complex way with the dynamical properties
(correlation and memory) of its links. Our model provides a novel
framework for systematic investigation of this delicate interplay\cg{,
  and for the description of real systems}: its simplicity allows for
efficient numerical simulation and analytical tractability, and its
flexibility allows us to explore and understand a wide range of
observable phenomena relating to diffusion over temporal networks.
Further to this it proves to be useful when investigating temporal
networks observed in the real world, where we cannot assume any
ability to study the effects of temporal correlations and memory in
isolation, thus making it an ideal building block for further studies
of empirical systems.

\section{ A general class of discrete autoregressive network models}
\label{darns}
Models for temporal networks in which links are governed by a possibly
correlated set of stochastic processes allow for a great deal of
control over various aspects of their output, but can run the risk of
being too abstract, and thus their use in describing empirical systems
can be limited.  For example, temporal networks in which links are
specified to have an inter-event time with a Weibull distribution have
been seen to reflect empirical findings with respect to infection
spreading, and clearly imply memory in the network, however it is not
clear that they are a good model for temporal networks in more general
settings \cite{Van_Mieghem_PRL13}.
\cg{Activity-driven network
  models \cite{perra2012activity}, in particular those versions with link
  reinforcement process \cite{karsai2014time}, allow us to describe
  non-Markovian memory in links. Nevertheless, it is unclear how to
  estimate activity-driven models on empirical data. State-space models of temporal
  networks, see for example
  \cite{hoff2002latent,sarkar2005dynamic,starnini2016model}, describe
  nodes as evolving in a latent Euclidean space and interacting
  depending on their `physical' distances in such a space.
  The resulting link dynamics
  can display both non-Markovian memory and cross-correlations among
  links. However, there is no explicit control on both
  features. Modelling temporal networks as Markov chains of generic
  memory order  \cite{peixoto2017modelling,peixoto2018change} allows us to characterise
  the memory patterns displayed by empirical data, however at the
  expense of high computational costs and the use of a large number of
  parameters.  Finally, maximum entropy models of temporal networks
  \cite{hanneke2010discrete} permit, in principle, to describe many
  patterns of link dynamics, also having a high level of control on the features
  of the output network, as well as allowing applications to empirical
  data.} \piero{Recently introduced Markovian models of temporal networks \cite{Williams_2019,mazzarisi2019dynamic} based on some opportune generalization of the Discrete AutoRegressive process \cite{jacobs1978discrete} are to all effects maximum entropy models, as shown in \cite{campajola2020equivalence}. Here, we show how the multivariate and non-Markovian generalization of the Discrete AutoRegressive mechanism \cite{jacobs1978discrete} is suited for a {\it general} description of the auto- and cross-correlation structure of temporal networks described as time series of adjacency matrices. Such a generalization allows us to define an entirely new class of models of temporal networks, which is highly flexible, largely controllable, and analytically tractable at the same time.}

\piero{The Discrete AutoRegressive process DAR(p) \cite{jacobs1978discrete}, whose properties have been largely studied in the statistics and econometrics literature \cite{jacobs1978discrete1,jacobs1978discrete2,jacobs1983stationary}, describes the persistence pattern of a stochastic process 
by means of the discrete autoregressive (copying) mechanism as}

\piero{
\be\label{darp}
 X_t = Q_tX_{t-Z_t}+ (1-Q_t)Y_t,
\ee}
\piero{with:  
\begin{enumerate}
\item $Q_t \sim \mathcal{B}(q)$ Bernoulli random variable with success
  probability $q$;
\item non-Markovian memory described by a random variable
  $Z_t$ which picks value $\tau$ running from $1$ to $p$ with
  probability $z_\tau$ (\ie {\it memory kernel}), such that
  $\sum_{\tau=1}^pz_\tau = 1$;
\item Bernoulli marginal $Y_t\sim\mathcal{B}(y)$ with success
  probability $y$.
\end{enumerate}}

\piero{The DAR(p) model in Eq.~(\ref{darp}) captures the {\it positive}
  autocorrelation of a binary time series with memory of generic order
  $p$ by means of the copying mechanism mediated by the Bernoulli
  random variable $Q_t$.
\\
  It is quite natural moving from the
  description of a single binary time series to the multivariate case
  of adjacency matrices, 
  thus exploiting the
  flexibility of the framework, to account also for the
  cross-interactions of links and time-varying patterns in temporal
  networks. For practical reasons, the multivariate generalization of the DAR(p)
  process in Eq.~(\ref{darp}) allows to define a new class
  of temporal networks, the so-called {\em Discrete
  AutoRegressive
  Network models}, which combine the mechanism of copying from the past
  with the sampling of links according to some marginal, which is
  Bernoulli in the simplest case. In particular, the latter
  can be interpreted to all
  effects as the non-Markovian dynamic generalization of the
  Erd\H{o}s-R\'enyi random graph model when one parameter is controlling for the density of the network.}

\piero{In order to precisely define the Discrete AutoRegressive
  Network models,  
let us consider a temporal adjacency matrix
  $\underline{\underline{A}}_t= \{ a^{ij}_{t} \}$, with
$t=1,2,\ldots$. If each link $(i,j)$ is labeled by a single index $\ell\equiv
  (i,j)$, we can consider the vectorization
  $\underline{\underline{X}}_t\equiv\{a^\ell_t\}^{\ell=1,...,L}$ of the
  adjacency matrix $\{a^{ij}_t\}^{(i,j)\in B}$ of the network snapshot
  at time $t$, where $L$ is the number of {\it possible} links
  belonging to some subset $B$ (\ie the so-called {\it backbone} of the
  temporal network) of all the $N(N-1)/2$ couples of nodes.
We then consider, 
  the following discrete autoregressive multivariate process ($\ell=1,\ldots,L$):}

\piero{\be\label{vdarp}
X_t^\ell = Q_t^\ell X_{t-Z_t^\ell}^{M_t^\ell} + (1-Q_t^\ell)Y_t^\ell
\ee}
\piero{with
\begin{enumerate}
\item $Q^{\ell}_t \sim \mathcal{B}(q^\ell_t)$ Bernoulli random variable with, in general, link-specific time-varying probability $q_t^\ell$;
\item non-Markovian memory described by a set of random variables 
  $Z_t^\ell$ which pick value $\tau$ running from $1$ to $p$ with
  probability $z_\tau$ (\ie {\it memory kernel}), such that
  $\sum_{\tau=1}^pz_\tau = 1$;
\item link cross-interactions described by a set of 
  random variables $M_t^\ell$ which pick values from $1$ to $L$ according
  to each row of a coupling matrix
  $\underline{\underline{C}}\equiv\{c^{\ell \ell'}\}$, 
  a row stochastic (\ie $\sum_{\ell'}c^{\ell\ell'}=1$) matrix, which
  characterises the correlations between pairs of links.
\item Bernoulli marginals $Y_t^\ell\sim\mathcal{B}(y^\ell_t)$ with, in
  general, link-specific time-varying probability $y^\ell_t$.
\end{enumerate}}

\piero{The formulation in Eq.~(\ref{vdarp}) is very general,
  accounting for time-dependent persistence patterns of links, with
  non-Markovian memory, cross-interactions mediated by the coupling
  matrix $\underline{\underline{C}}$, and possibly time-varying
  marginal probabilities. In practice, some parametrization needs to
  be considered, and it is possible to reduce the complexity of the
  model and use it to focus, one by one, on the various specific
  features of temporal networks.} 

\piero{Previous works have started to investigate the role of
  non-Markovian memory in models of temporal networks that can now be
  seen as extreme limiting cases of the most general framework
  proposed in Eq.~(\ref{vdarp}). However, a very imporant
  aspect, which has not yet received the deserved attention is the
  modelling of cross-interactions of links.} 
\piero{For instance, the dynamics of spreading processes on temporal networks
  with memory has been investigated in the DARN(p) model, 
a non-Markovian
model that can be seen as a limiting case of the model
in Eq.~(\ref{vdarp}) with constant parameters and with
diagonal coupling matrix $\underline{\underline{C}}$, i.e. under 
the very strong assumption that only auto-correlations in the link
dynamics are present \cite{Williams_2019}.}
\piero{The authors of Ref. \cite{mazzarisi2019dynamic} have instead
  proposed an empirical application of link inference to 
  the interbank market. The have considered a model similar to that
  in  Eq.~(\ref{vdarp}) with Markovian link-persistence patterns,
  again without explicit cross-correlations of links, but combined
  with node-specific time-varying marginals.} 
\piero{Finally, Granger causality has been investigated in a model
  with two time series, which corresponds to a {\it bivariate}
  case (\ie $\ell=1,2$) of the model in
  Eq.~(\ref{vdarp}) with constant parameters 
  \cite{mazzarisi2020tail}. }

\piero{Here, we focus on a crucial feature of real-world temporal
  networks that has received less attention from a modelling point of
  view, namely the cross-interactions of links (i.e. the presence of
  dependencies in the time evolution of pairs of different links). 
In particular, we will consider the model in Eq.~(\ref{vdarp})
with general memory kernels and opportune parametrizations of the
coupling matrix $\underline{\underline{C}}$. In this way we will be
able to study auto-
  and cross-correlated link dynamics combined with non-Markovian
  memory, and in the presence of a backbone network defining which links
  may be present or not.}
  
\piero{We point out the richness, versatility 
  and controllability of our model of temporal networks, together with
  its low computational costs (thanks to maximum
  likelihood methods for inference) in empirical applications.
  In Section \ref{section2}, we will first consider the CDARN($p$) model,
  a simplified version of the model
  in Eq.~(\ref{vdarp}) with constant parameters $q_t^\ell=q$
  and $y_t^\ell=y$ $\forall t,\ell$. Then, in Section \ref{empcdarn}, 
  thanks to the high flexibility of
  the proposed framework, we will relax the last assumption by allowing for
  heterogeneous parameters (\ie link-specific $q^\ell$ and $y^\ell$) and we will 
  exploit {\it local} likelihood methods \cite{hastie2009elements} to
  deal with time-varying parameters $q_t$, $c_t$, and $y_t$. Finally, the 
  analytical tractability of our approach will come to light in Section
  \ref{diff_process}  in the study
  of the dynamics of spreading processes over temporal networks
  generated by the model.}


\section{The CDARN($p$) model of temporal networks}
\label{section2}

Here, we consider a simplified version of the general model of temporal networks in Eq.~(\ref{vdarp}), which is rich enough to describe non-Markovian
  memory patterns, with precisely controlled strength and length of
  the memory, while also reproducing a key feature of real-world
  networks, namely correlations between the evolutions of links over
  time, as produced by dependencies between their dynamics.
\cg{Furthermore, we want to keep such a model as simple as possible,
    with a small number of parameters, thus permitting also easy
    application to empirical data by fitting the parameters of the model
    on graph sequences from the real world.}
\piero{Hence, we take the general setting given by Eq.~(\ref{vdarp}) and consider a particular parametrization that reflects} \cg{ the presence 
of two key features of real systems \cite{delvenne2015diffusion,Fallani08,Yoneki:2009,MIN2011606,Valencia08}:}
\cg{1) the existence of an underlying restriction, a so-called
  network ``backbone'' on which links can occur; ~ 
2) the presence of cross-correlations in link dynamics, \ie of
  dynamical dependencies in the temporal activities of different
links.}
\\
\cg{The model we introduce, the so-called 
{\em Correlated Discrete Auto-Regressive Network} model of order
$p$, or in short the CDARN($p$) model, describes the dynamics of links with an included
  mechanism for copying from the past: at each time, a link (or
  no-link) can be copied from the past, either of the link itself or the
  past of some other link on the backbone, or sampled according to a
  Bernoulli marginal (Erd\"{o}s-R\'{e}nyi model). Which point in the past, from $1$
  to $p$ steps behind, is then randomly selected with uniform
  probability. The model is, in effect, the \piero{multivariate generalisation of the DARN(p) model \cite{Williams_2019} with non-Markovian memory and both self-
  and cross- interactions of links. In the following, for clarity, we briefly review DARN(p) before}}\cg{\piero{introducing the CDARN(p) model}, while in the
next section we show how to estimate the model on real data, to infer
the key parameters of the network dynamics. Once estimated on data,
the model can be used also for link prediction. Moreover in Section
\ref{diff_process} we will show how the model parameters can be varied to
study in a systematic and controlled way the role that memory in the
underlying temporal network has on the rate at which information, or
some other quantity spreads throughout a system whose interactions
change over time.}

\subsection{The basic model}
\label{cdarn_description}

The DARN($p$) model, originally introduced in \cite{Williams_2019}, generates
a temporal network with precisely controlled memory features in the temporal
sequence of each link.  
Namely, the model considers $N$ nodes and  
assigns to each of the $N(N-1)/2$ pairs of nodes 
the presence or absence of a link as ruled by 
independent, identical DAR($p$) processes 
(Discrete Auto-Regressive processes of order $p$) 
\cite{mazzarisi2019dynamic,jacobs1978discrete,hamilton1995time,Runge:2015aa}.
 In this way each link will, at each time
step, either be generated randomly with some fixed probability, or will copy 
a randomly chosen state from its past $p$ iterations.
In terms of random variables, this gives us a 
temporal adjacency matrix $ \underline{\underline{A}}_t= \{ a^{ij}_{t} \}$, with $t=1,2,\ldots$,
where each link $(i,j)$,
with $i,j=1,\ldots,N$ is governed by the process: 
\begin{equation}
	a^{ij}_t = Q^{ij}_t a^{ij}_{(t-Z^{ij}_t )} + (1-Q^{ij}_t)Y^{ij}_t.
    \label{DARp_eqn}
\end{equation}
where, for each link $(i,j)$ and time $t$, $Q^{ij}_t$, $Y^{ij}_t$
  and $Z^{ij}_t$ are random variables. In particular,  
$Q^{ij}_t \sim \mathcal{B}(q)$ and
  $Y^{ij}_t \sim \mathcal{B}(y)$ are Bernoulli random variables,
  while $Z^{ij}_t$ is some random variable which picks integers in the range $\{1,...,p\}$. \piero{Note that no restrictions are imposed to the memory kernel controlling for the probability of picking the integers in the range $\{1,...,p\}$, as long as the probability sums to one. For example, an uniform kernel describes equal probabilities of picking past observations, from lag $1$ to lag $p$, whereas an exponential kernel describes probabilities exponentially decaying to zero as the lag is increasing. }
  Without loss of generality, here we take
$Z^{ij}_t \sim Uniform(1,p)$.
The networks created by the DARN($p$) model are undirected, and clearly 
non-Markovian, with precise memory $p$.

The DARN($p$) model assumes that links
can occur between any two nodes. This is not 
always the case in real world networks, where certain links may
be unfeasible, or simply impossible. For example, a plane may 
not be allowed to fly between two particular airports, or a 
doctor may be responsible for a small number of patients, and
therefore not interact with others. We therefore say that these 
temporal networks have a ``backbone": a fixed set of possible 
links which restrict the networks evolution. With this in mind 
we make our first modification leading to a more general framework. First, 
we define a {\em backbone network} with $L$ links described
by a static $N \times N$ adjacency matrix $ \underline{\underline{B}}=\{ b^{ij} \}$.
Then a temporal network on this backbone is represented by a $N
\times N$ time-varying adjacency matrix $ \underline{\underline{A}}_t = \{ a_t^{ij} \}$, so
that $a_t^{ij} = 0$ for all $t$ if $b^{ij} = 0$, while if $b^{ij}=1$
then the link $(i,j)$ can exist for any value of $t$.
%
In this way the presence of links can be appropriately limited
to reflect reality.

Since links in the DARN($p$) model are generated by independent
processes, there can be auto-correlations in the temporal activity
of each link, but no cross-correlations between different links.
Conversely, correlations
among different links are a natural feature of many systems.  To
further our earlier analogy, an airline is unlikely to schedule two
flights between the same airports in close proximity to each other,
but may prefer to schedule flights at appropriate times to make
connections. Similarly doctors may see patients in a particular order
each day, even if the duration of each interaction is not so
consistent. In order to allow for such correlations, we introduce our
second modification: when a link in a DARN($p$) model would pick from
its own memory, we now allow it to pick a link from the network at
random, possibly itself again, and copy a randomly chosen state of
that link instead. In this way, the dynamics of each link $(i,j)$
that belongs to the network backbone is governed by the process:
\begin{equation}
	a_t^{ij} = Q^{ij}_t a_{(t-Z^{ij}_t )}^{M^{ij}_t} + (1-Q^{ij}_t)Y^{ij}_t 
	\label{CDARN_rv_eqn}
\end{equation}
with $i,j=1,\ldots,N$ and such that $b^{ij}=1$, and where  
at each time $t$, $M^{ij}_t$ is a (categorical) random variable which  
associates to link $(i,j)$  
another link $(i',j')$ among
links which are present in the backbone $ \underline{\underline{B}}$, with an
assigned probability distribution. Note that for each time $t$ and link $(i,j)$,  
$M^{ij}_t$ is independent
and identically distributed. That is to say, if a link is copied from
the past of another link, then which link it chooses is completely independent on
either the time, or the existence of any other link.

\cg{Hence, the CDARN($p$) model in Eq.~(\ref{CDARN_rv_eqn}) relies on the
  following input parameters. The first ingredient is the $N \times N$
  adjacency matrix $ \underline{\underline{B}}$ describing the
  structure of the underlying network backbone of $N$ nodes and $L$
  links, i.e. defining, which pairs of nodes can be connected by links
  and which pairs cannot. The backbone has density $D=2L/N(N-1)$, if
  the network is undirected.  However at each time not all the links
  of the backbone are necessarily present, and the average link
  density within the backbone is controlled by the parameter
  $0<y<1$.} Moreover $0 \le q \le 1$ and $p = 1, 2, \ldots$ are
  respectively the strength and length of the memory component of the
  dynamics of the temporal network. 
%
\cg{Finally, the structure of interactions among links is captured by
 the random variable $M^{ij}_t$, which can be described by a $L
  \times L$ link coupling matrix $ \underline{\underline{C}}$,
  characterising the correlations between pairs of links.}
Labelling links with a linear index $(i,j) \mapsto \ell$, $(i',j')
\mapsto \ell'$, with $\ell,\ell'=1,2,\ldots,L$ (see Appendix A for a
full explanation), then $M^{ij}_{t}$ can be characterised by the
probabilities:
$$\text{Prob}(\ell \text{ draws from } \ell' ) = c^{\ell \ell'}$$
These probabilities define a $L \times L$ 
row-stochastic matrix $ \underline{\underline{C}} =\{  c^{\ell \ell'} \}$,
which we call the {\em coupling matrix}. By tuning the entries of this matrix 
we can specify 
the dependencies among links existing in our temporal network. 
In practice, for each possible link
  $(i,j)$ and at each time $t$, $M^{ij}_t$ will select another link 
  $(i',j')$ among a set of possible links associated to $(i,j)$, 
  as given by matrix $ \underline{\underline{C}}$. Then, the presence of the term
  $a_{(t-Z^{ij}_t )}^{M^{ij}_t}$ in Eq.~(\ref{CDARN_rv_eqn}),
  represents the state of link $(i',j')$ at one of the previous $p$
 temporal steps, 
and so will allow link $(i,j)$ to copy its state at time $t$
from one of the $p$ past states of link $(i',j')$.
\cg{This is similar to building the line graph associated with 
the original network, but in the temporal case and restricting to pairs of links 
which are on the backbone.}
The choice of the coupling matrix $\underline{\underline{C}}$ is
\cg{a crucial}
part of the CDARN($p$) model,
\cg{
as this defines which links dynamics are correlated. Since the matrix has a large number of entries, 
it is advisable to choose a parsimonious representation depending on a small number of parameters. 
There are many ways one could structure 
 the matrix $ \underline{\underline{C}}$, which we refer to as ``coupling models".}
 Here, we will focus on the following three simple approaches: 
 (i) only link autocorrelations but no cross correlations between
 different links, (ii) links are coupled to all other 
 neighbouring links in the network backbone (as defined by $ \underline{\underline{B}}$)
 with equal strength, (iii) links are coupled to all other links 
 in the backbone with equal strength.

To summarise, given a backbone $ \underline{\underline{B}}$ with $L$ links, we have the
three following coupling models:
\begin{enumerate}
\item The {\em no cross correlation (NCC)} coupling model, where the coupling
  matrix reads  $ \underline{\underline{C}} =  \underline{\underline{I_d}}$ (the identity matrix).
\item The {\em local cross correlation (LCC)} coupling model, where
  the entries of the coupling matrix can be written as: 
$c^{\ell \ell'} = (1-c) \delta(\ell, \ell') + \chi(\ell' \in \partial_{B} \ell) \, {c}/{\left| \partial_{B} \ell \right|}$,
for coupling strength $c$.
  Here, $\chi$ is the indicator function, $\delta(\ell, \ell') = 1$ if $\ell=\ell'$ and
  0 othwerwise,  and 
$\partial_{B} \ell$ is the neighbourhood of link $\ell$ in backbone $ \underline{\underline{B}}$, i.e for $\ell = (i,j)$ 
$\partial_B \ell = \{\ell' = (i',j') : b^{i'j'} = 1 \text{ and }  i' \in \ell \text{ or } j' \in \ell\}$.

\item The {\em uniform cross correlation (UCC)} coupling model, where the entries
of the coupling matrix can be written as $c^{\ell \ell'} = (1-c)\delta(\ell, \ell')  + (1- \delta(\ell, \ell')) c/(L-1)$, for coupling strength $c$.
\end{enumerate}
Notice that the parameter $0 \le c \le 1$ in the second
and third coupling model
allows us to tune the contribution of the cross correlations with respect to
that of the autocorrelations. In particular the NCC model is the 
special case of the LCC and UCC models with coupling strength $c = 0$.

Considering all the building blocks, we then have our full model, which we name the {\em Correlated
  Discrete Auto-Regressive Network} model of order $p$, or in short
CDARN($p$) model.  This model has the advantage of being able to
introduce both auto- and cross-correlations in the link activities in
a controlled way, allowing for a more realistic description of real
world systems. It also retains a lot of the simplicity and
tractability of the DARN($p$) model. Indeed, \piero{three} of the key features
of the DARN($p$) model that allow us to study a range of phenomena are
exactly the same.
\cg{Namely, the (unconditional) probability of observing a link (restricted to the feasible connections over the backbone) is $y$, similarly to ER graphs (See Appendix B) \cite{bollobas2001random}.}
Moreover, in the limit of long memory, as
given by large $p$, the model is identical to a sequence of
uncorrelated ER graphs (see Appendix C). \piero{Finally, the inter-event time distribution, also known as inter-contact time in human communication networks, is ({\it approximate}) exponential, with a time scale that is bounded from above by the DARN(p) model. (see Appendix \ref{cdarn_validation})}

In summary, our model generates temporal networks $ \underline{\underline{A}}_t,
t=1,2,\ldots$, with precisely controlled coupling among links, given
the following set of control parameters: network backbone as specified
by matrix $ \underline{\underline{B}}$, link density $y$, memory strength $q$, memory length
$p$, and link coupling matrix $ \underline{\underline{C}}$. For our purposes we will assume
that the links in the temporal network are undirected, we do this by
identifying $a^{ij}_t = a^{ji}_t$. Implicitly the backbone in any
network will be taken as undirected, implying that only
symmetric matrices $ \underline{\underline{B}}$ will be considered. 
The extension to directed networks is, however, straightforward.
Finally, an important
assumption of the CDARN($p$) model is that the parameter $p$, $q$, and
$y$ are the same for all the links of the backbone. Of course this
choice is a simplification only motivated by the need for control over
the dynamics with a small number of parameters, \piero{for both empirical application to real-world networks and analytical study of the dynamics of spreading processes over temporal networks}. The CDARN($p$) model 
can in fact be easily generalised to the case of \piero{link-specific or time-varying parameters}, as done \piero{in the section below or}, for instance, in the simpler DARN($1$) model 
presented in \cite{mazzarisi2019dynamic}.

\subsection{Possible generalision of the model}
\label{cdarn_gen}

\piero{Real-world networked systems can be characterized by heterogenous, \ie link- or node-specific, and/or time-varying patterns for link dynamics. Realistic models of temporal networks need to be able to capture such patterns when we aim to replicate empirically observed network dynamics. Such a generalization can easily be accounted for within our framework by considering link-specific parameters $y\rightarrow y^\ell$ and $q\rightarrow q^\ell$, or promoting constant parameters $q$, $c$, and $y$ to time-varying parameters $q_t$, $c_t$, and $y_t$, then introducing a method to estimate them.}

\piero{Below, we generalize the simplified version of the CDARN(p) model by relaxing, step by step, some homogeneity assumptions:
\begin{enumerate}
\item ceteribus paribus, the probability of success $y$ of the Bernoulli marginal is not anymore equal for all links, but each link $\ell$ is described by a different probability $y^\ell$, thus allowing for densities that change link by link;
\item ceteribus paribus, the probability of copying from the past $q$ is not anymore equal for all inks, but each link  $\ell$ is more or less persistent depending on a specific parameter $q^\ell$;
\item parameters $q$, $c$, and $y$ are not constant anymore during the evolution of the network, but they can change in time, in order to capture the presence of time-varying, possibly non-stationary, patterns in link dynamics, \eg link density and/or correlations depending on the time of the day.
\end{enumerate}
}

\section{Network model inference and link prediction}
\label{empcdarn}

\cg{In this section we present a method for estimating the model parameters
of the CDARN(p) from real data.  To this end, the strength of
our approach comes to light, since the CDARN(p) model defined in
Eq.~(\ref{CDARN_rv_eqn}) can be estimated on data by using maximum
likelihood methods \cite{hastie2009elements}, thus inferring case by
case the role played in real world by both auto- and cross-
correlations of links. \piero{Then, thanks to the flexibility of maximum likelihood approach for inference, we show that our methodology can easily accomodate for {\it heterogeneous} or {\it time-varying} parameters, thus better capturing the dynamics of real-world networked systems.} \piero{Last but not least, we prove empirically that} the inclusion of
cross-interactions in the description of the link dynamics is no small
matter: cross-interactions of links are essential in describing various
types of real-world temporal networks. In particular, we show, through a link
prediction study, that such correlation patterns are, indeed, present
in networks from the real world.}

\subsection{\cg{Parameter estimation}}

\cg{Assume we have observed a time series of network snapshots
$\{a^{ij}_t\}_{t=p+1,...,T}^{i,j=1,...,N}$ with given initial $p$
conditions $\{a_t^{ij}\}_{t=1,...,p}^{i,j=1,...,N}$, then we ask for
the values of parameters in Eq.~(\ref{CDARN_rv_eqn}) which best
describe the evolution of the temporal network.} \piero{Here, we aim to obtain a {\it point estimate} of the parameters, which is, from a Bayesian inference perspective, the value maximizing the posterior probability of parameters given the data.}
\piero{By referring to the set of parameters as $\theta$, thanks to the 
Bayes theorem, we can write: 
$$
\mathbb{P}(\theta\vert \underline{\underline{A}})\propto \mathbb{P}(\underline{\underline{A}}\vert \theta)\mathbb{P}(\theta)
$$
where $\underline{\underline{A}}\equiv\{\underline{\underline{A}}_t\}_{t=1,\ldots,T}$. Without prior information on the parameters, we can assume uniform prior distribution $\mathbb{P}(\theta)$. Thus, the point estimation corresponds to maximizing the likelihood of data under the model with parameters $\theta$, \ie $\mathbb{P}(\underline{\underline{A}}\vert \theta)$,}
\cg{ namely
the Maximum Likelihood Estimator (MLE) of the CDARN(p) model.  The
{\it likelihood} of data \piero{under the CDARN(p) model} reads as
\begin{equation}\label{lcdarn}
\begin{split}
\mathbb{P}(\underline{\underline{A}}_{p+1},...,\underline{\underline{A}}_T\vert & \underline{\underline{A}}_1,...,\underline{\underline{A}}_p,q,c,y) \\
	&=\prod_{t=p+1}^T \mathbb{P}(\underline{\underline{A}}_t\vert \underline{\underline{A}}_{t-1},...,\underline{\underline{A}}_{t-p},q,c,y),
\end{split}
\end{equation}
by using the Markov property, and with $\{q,c,y\}$ the model parameters.
The likelihood of the Markov chain then corresponds to the product of $T-p$
conditionally independent transition probabilities, each one
describing the likelihood of a network snapshot given the previous $p$
observations, because of the non-Markovian memory of the process. The
MLE of the parameters is thus obtained by maximising
Eq.~(\ref{lcdarn}), or equivalently the {\it log-likelihood}
$\mathbb{L}(q,c,y)\equiv \log
\mathbb{P}(\{\underline{\underline{A}}_{t}\}_{t=p+1,...,T}\vert
\{\underline{\underline{A}}_\tau\}_{\tau=1,...,p},q,c,y)$, that is
\begin{equation}\label{mlecdarn}
\arg\max_{y,c,q} \sum_{t=p+1}^T\log \mathbb{P}(\underline{\underline{A}}_t\vert \underline{\underline{A}}_{t-1},...,\underline{\underline{A}}_{t-p},q,c,y),\:\:\:q,c,y\in[0,1].
\end{equation}
The solution to Eq.~(\ref{mlecdarn}) is the MLE
$\{\hat{q},\hat{c},\hat{y}\}$ of the CDARN(p) model. The explicit
formulas for the MLE are in the Appendix Section \ref{sm_mle}. Notice
that the order $p$ of the memory of the Markov chain can be selected
by finding the integer value which maximises the likelihood of data \piero{under CDARN(p)},
since the number of parameters of the CDARN(p) model is the same,
independently from the order $p$ and thus there is no need to
penalise the use of more parameters. See \cite{williams2020shape}
for a study on the optimal selection of the (non-Markovian) memory in
temporal networks.}

\cg{In empirical applications, some networked systems may display some
  time-varying density pattern, related for example to the activity of
  nodes, which may, crucially, affect the estimation of the parameters
  $q$ and $c$. For example, in the presence of a seasonality pattern,
  \ie a network density depending on the time of the day, considering
  a constant density parameter tends to overestimate correlations,
  thus the MLE $\hat{q}$ and $\hat{c}$. In our framework, any
  variation of network density can be taken into account by letting
  $y$ become a time-varying parameter, $y\rightarrow y_t$, and using
  the (suboptimal) estimator $\hat{y}_t=L^{-1}\sum_{(i,j)\in B}
  a_t^{ij}$.  Hence, when network density is clearly not constant, a
  two-step estimation procedure can be implemented. First, we
  estimate the time series of density parameters
  $\{\hat{y}_t\}_{t=p+1,...,T}$.  Second, the MLE $\hat{q}$ and
  $\hat{c}$ are obtained by solving Eq.~(\ref{mlecdarn}), but
  conditioning on the values $\{\hat{y}_t\}_{t=p+1,...,T}$.
In the presence of some density pattern, we use this method to obtain a genuine estimation of the memory parameters.}

\subsection{\piero{Heterogenous and time-varying parameters}}


\piero{In the case of heterogenous parameters, the MLE problem in Eq.~(\ref{mlecdarn}) can be generalized and solved, similarly to what has been done in \cite{mazzarisi2019dynamic}. We study explicitly such MLE problem in the appendix section \ref{sm_hetero_mle}.}

\piero{In the case of time-varying parameters, we can use a nonparametric technique based on {\it local} likelihood estimation to infer the dynamics of $q_t$, $c_t$, and $y_t$. The main idea relies on considering observation weights, which are decaying in time, in the maximum likelihood equations. Thus, to obtain a local (in time) estimate of parameters at time $t$, we fit the model by using those observations that are closer to the time snapshot $t$. This localization is achieved via a weighting function or {\it kernel} $K_\lambda(t,s)$ with {\it bandwidth} $\lambda$, which assigns a weight to $s$ based on the time difference $\vert s-t\vert$. Here, we use the {\it Epanechnikov} quadratic kernel}
\piero{
$$
K_\lambda(t,s) = \begin{cases}
\frac{3}{4}\left[1-\left(\frac{\vert s-t\vert}{\lambda}\right)^2\right]\:&\mbox{if}\:\:\frac{\vert s-t\vert}{\lambda}<1\\
0\:&\:\mbox{otherwise}
\end{cases}
$$
with $\lambda=40$.} \piero{For further details on the local likelihood method see Ref.~\cite{hastie2009elements}, while for a study on the optimization of the bandwidth of the kernel see Ref.~\cite{fan1998local}.}

\piero{Hence, the local MLE problem at time $t$ reads as
\begin{equation}\label{localmlecdarn}
\arg\max_{y_t,c_t,q_t} \sum_{s=p+1}^T K_\lambda(t,s) \log \mathbb{P}(\underline{\underline{A}}_s\vert \underline{\underline{A}}_{s-1},...,\underline{\underline{A}}_{s-p},q_t,c_t,y_t),
\end{equation}
with $q_t,c_t,y_t\in[0,1]$. The maximum likelihood equations to solve follow as similar to the standard ones. Then, by rolling the kernel over time $t$, a nonparametric reconstruction of the dynamics of time-varying parameters is obtained.}

\subsection{\cg{Application to real temporal networks}}

\cg{In this empirical Section, we consider the application of the CDARN(p) model to temporal networks, described as time series of adjacency matrices, each one capturing the links between the nodes of the network, within a given time resolution. Each link describes a particular interaction, typical of the networked system under investigation. The following data sets are considered:
\begin{enumerate}
\item transportation networks, \ie bus (B), (underground) rail (R), and train (T), designed for the public transport in Berlin (B), Dublin (D), Helsinki (H), Paris (P), Rome (R), Sydney (S), Venice (V), Winnipeg (W), namely records for the movements of public transport systems from stop to stop\footnote{Here, the number of nodes is corresponding to the number of stations, or stops, characterizing the specific public transport at each city.}, with time resolution of 1 minute. In particular, a connection between two stops is associated with a time interval, from the departure to the arrival, thus, within our framework, a link appears at the network snapshot corresponding to departure, then lasts up to the snapshot which includes the arrival;
\item online social communication networks, in particular
\begin{enumerate}
\item text message interactions between 101 college students (MSG), namely messages sent between (anonymised) student users of an online communication platform at the University of California, Irvine, over a period of seven months, with a time resolution of one hour;
\item email communications (EM), namely internal e-mail communications between 65 employees of a mid-sized manufacturing company over a period of nine months, with time resolutions of 5, 10, 30 minutes, 1 hour, and 24 hours;
\end{enumerate}
in both cases, a link is an instantaneous communication between two nodes and it is described by the entry of the adjacency matrix associated with the network snapshot of all communications within the considered time window;
\item social interaction or contact network (CN), which can be seen also as a off-line social communication network, namely the interactions (measured by bluetooth devices - phones - carried by) of 94 students at MIT over eight months, with time resolutions of 10, 30, 60, 120 minutes. Here, a link is a contact between two nodes, lasting for the time of the interaction, measured as the number of network snapshots at which the link is present;
\item football networks, \ie the temporal networks formed by footballers (F) over a match, for two different
games (1,2), and for both sides separately (home h, away a), with time resolutions of 1, 2, 10, 30, and 60 seconds. A link is a contact between two players, measured as a physical distance between each other below a threshold of 10 meters. Thus, a link is described by an entry of the adjacency matrix associated with the network snapshot at which the contact occurs. Then, a link lasts for all the time snapshots at which the contact is present.
\end{enumerate} 
}

\cg{Finally, the backbone is built by considering the network of all
  pairs connected at least once in the whole time period, for each
  temporal network.}

\begin{figure*}
\centering
\includegraphics[width=0.99\textwidth]{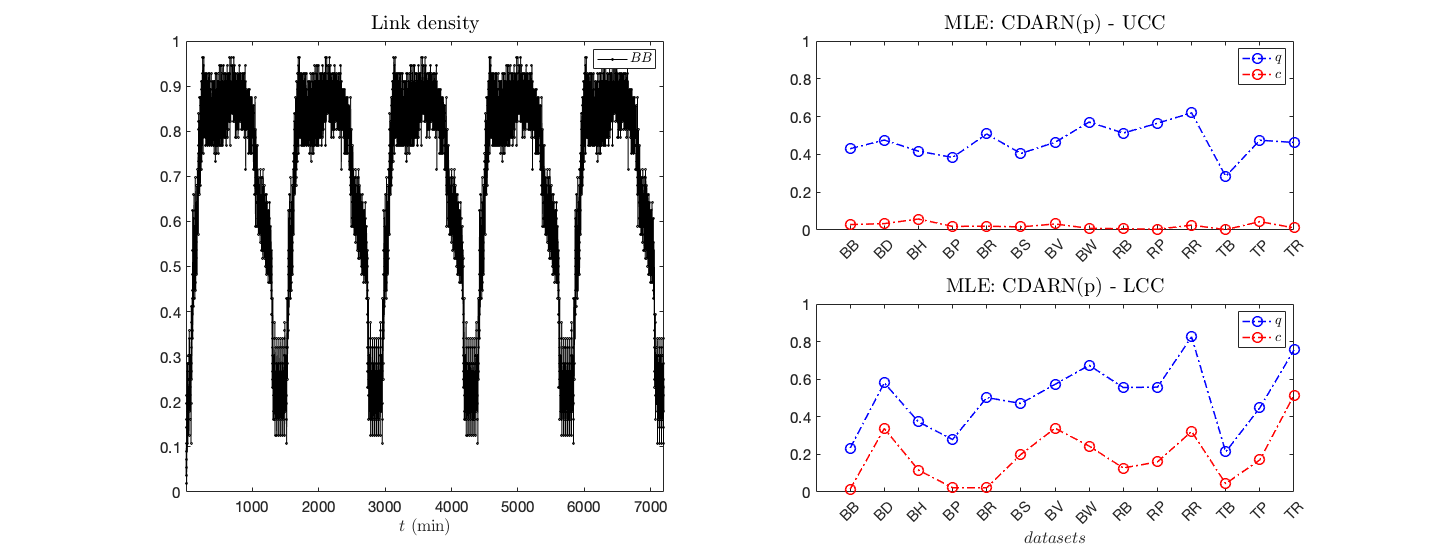}
\caption{\cg{{\bf Link density as a function of time (left) and MLE of the parameters $q$ and $c$} of the CDARN(p) model (right), considering both UCC (top right) and LCC (bottom right) coupling models, for the transportation network datasets (as described in the main text).}}
\label{figTransport}
\end{figure*}

\begin{figure*}
\centering
\includegraphics[width=0.99\textwidth]{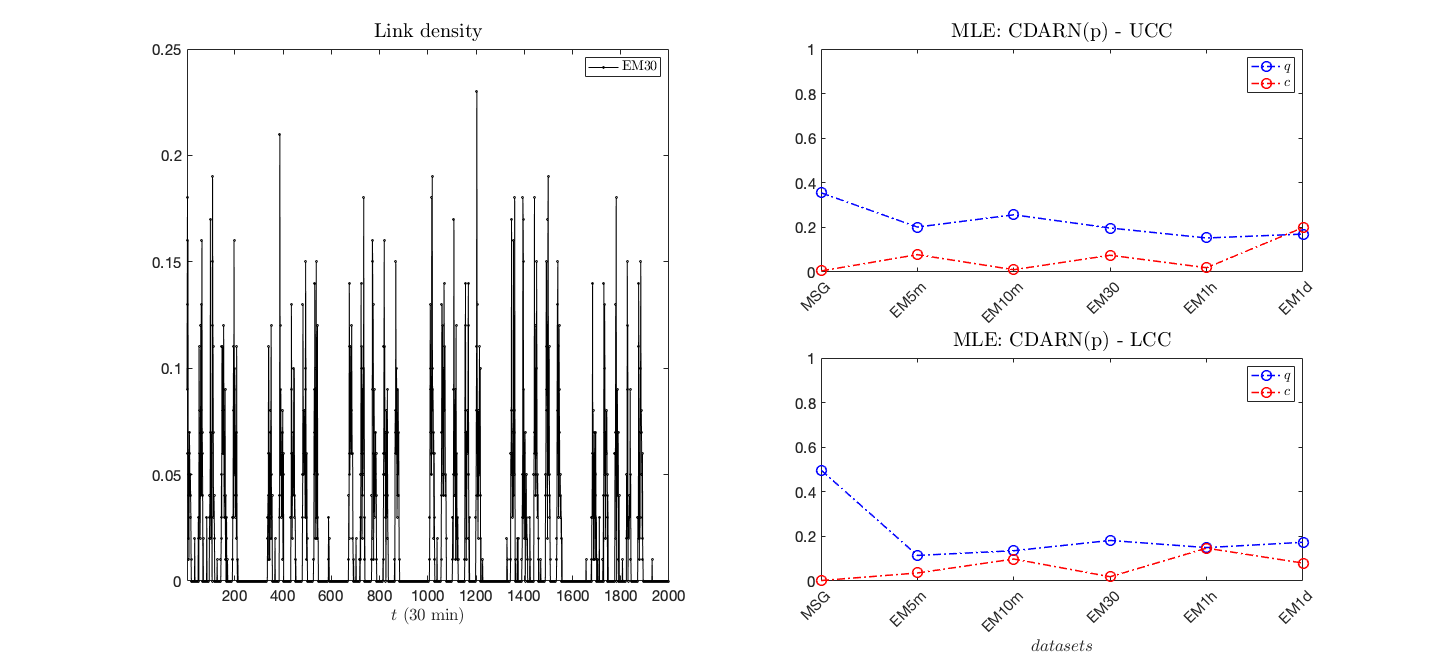}
\caption{\cg{{\bf Link density as a function of time (left) and MLE of the parameters $q$ and $c$} of the CDARN(p) model (right), considering both UCC (top right) and LCC (bottom right) coupling models, for the online social communication network datasets (as described in the main text).}}
\label{figEmails}
\end{figure*}

\begin{figure*}
\centering
\includegraphics[width=0.99\textwidth]{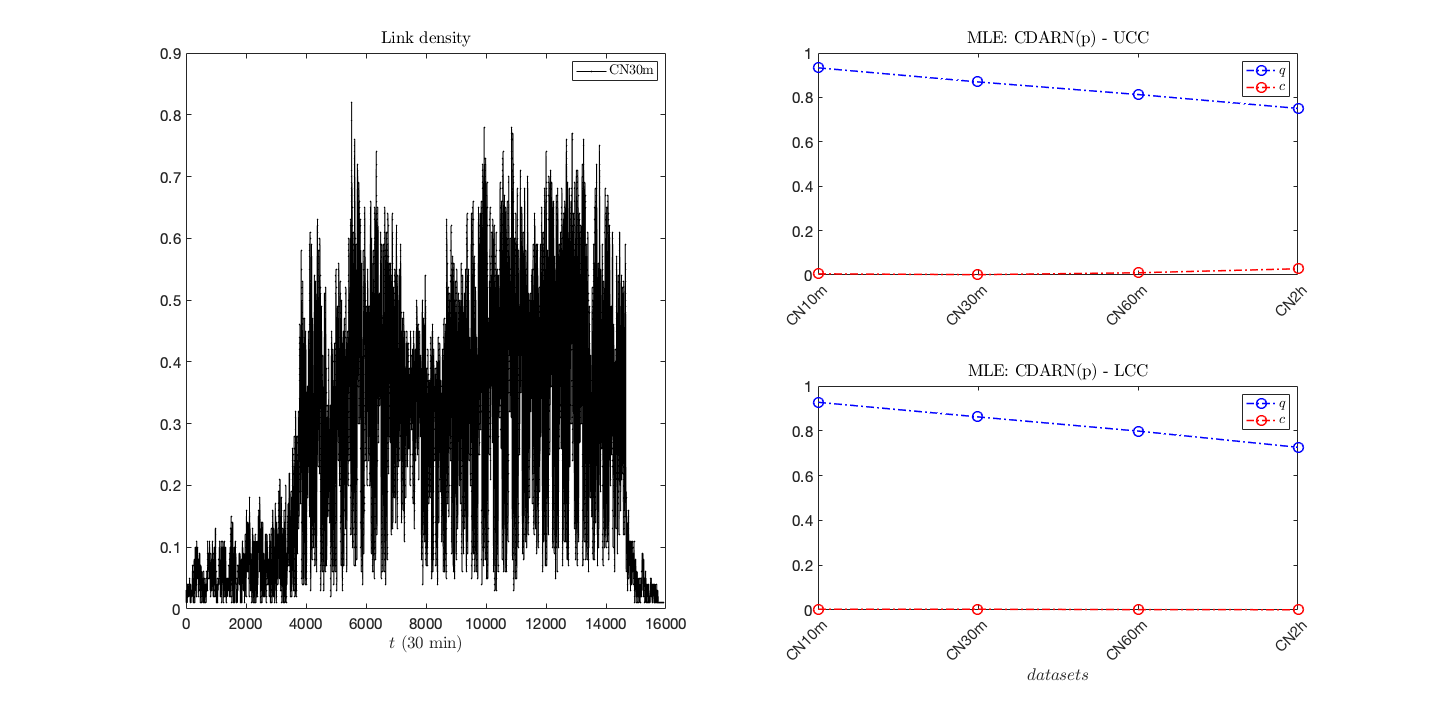}
\caption{\cg{{\bf Link density as a function of time (left) and MLE of the parameters $q$ and $c$} of the CDARN(p) model (right), considering both UCC (top right) and LCC (bottom right) coupling models, for the contact network datasets (as described in the main text).}}
\label{figContact}
\end{figure*}

\begin{figure*}
\centering
\includegraphics[width=0.99\textwidth]{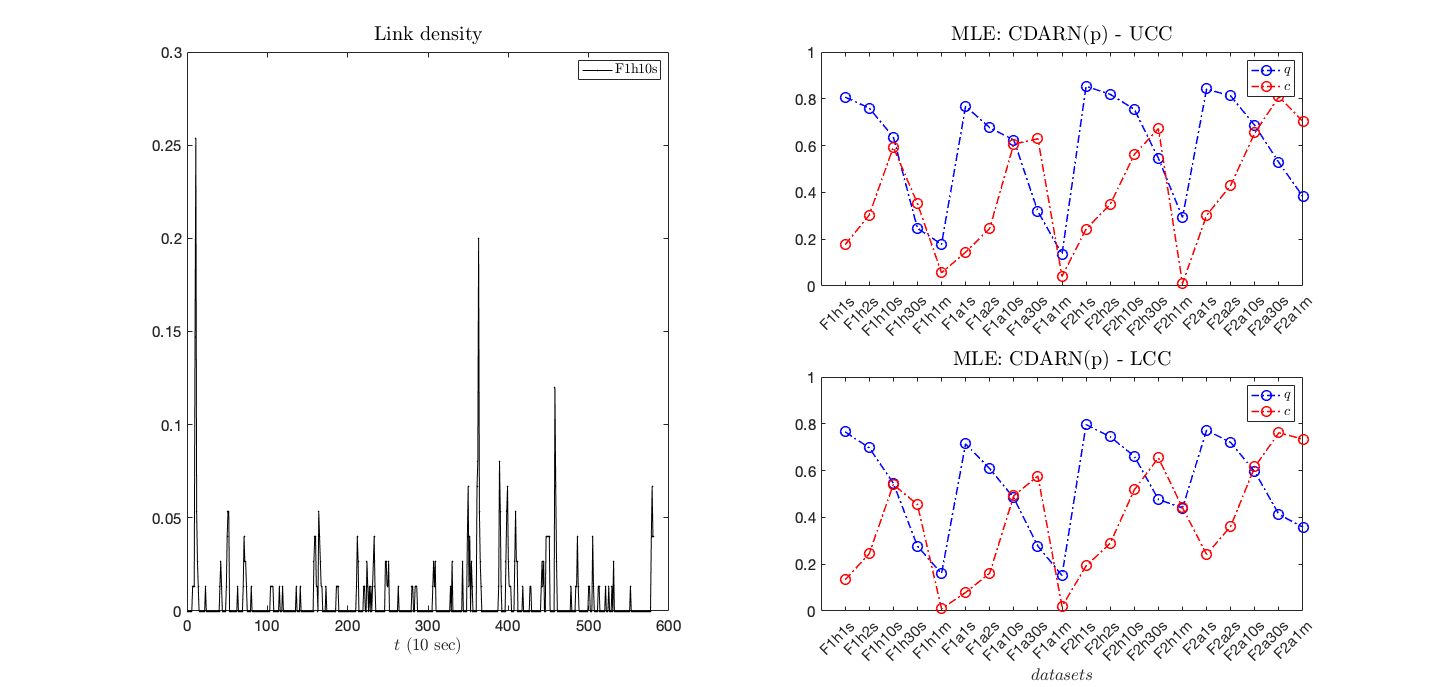}
\caption{\cg{{\bf Link density as a function of time (left) and MLE of the parameters $q$ and $c$} of the CDARN(p) model (right), considering both UCC (top right) and LCC (bottom right) coupling models, for the football network datasets (as described in the main text).}}
\label{figCalcio}
\end{figure*}

\cg{We then estimate the CDARN(p) model on these network datasets, by
considering the correction for the seasonality or non-stationarity
patterns displayed by link density for the first three types of
temporal networks, see the left panels of Figs.~\ref{figTransport},
\ref{figEmails}, and \ref{figContact}, while no correction is applied
to the football networks, as supported by empirical evidence (see the
left panel of Fig.~\ref{figCalcio}), but we have nevertheless solved the original
problem stated in equation (\ref{mlecdarn}). The maximum likelihood estimators of the
parameters $q$ and $c$ are shown in the right panels of Figs.~\ref{figTransport}, \ref{figEmails}, \ref{figContact}, and
\ref{figCalcio}, for both the Uniform Cross Correlation (UCC) and
Local Cross Correlation (LCC) coupling models. In both cases, the
order $p$ is selected by maximising the
likelihood of observing the given data under the CDARN(p) model.}

\cg{
\begin{enumerate}
\item For transportation networks, the mechanism of copying from the past captures the observed link persistence patterns (long-lasting connections between two stops) as well as cross interactions (transport connections at the stop), as testified by the large values of $q$. However, cross interactions become significant only restricting to neighbour links over the backbone, see the estimated values of $c$ (red points) in the right panels of Fig.~\ref{figTransport}, which are close to zero for UCC, significantly larger than zero for LCC (for almost all datasets). This behaviour is consistent with the underlying dynamics of the considered transportation systems, where a link is a connection between two physical stops, and interactions may arise only between incoming or outgoing transport connections at the same stops. Furthermore, we can notice a significant positive correlation between the estimated parameters $q$ and $c$ in this case. Finally, the order $p$ of the CDARN(p) model is estimated as one (generating a Markovian network), for all types of transportation and for all cities.
\item Online social networks display less important memory patterns, as testified by small values of $q$ and $c$, see the right panels of Fig.~\ref{figEmails}. Here, differently from above, we do not notice much difference between UCC and LCC coupling models. However, non-Markovian effects characterise such networks. In fact, for the UCC model we obtain $p=4$ for the MSG network, $p=5$ for the EM dataset with $5$ minutes resolution, $p=4$ for the EM with $10$ minute resolution, $p=2$ for the EM with $30$ minute resolution, $p=1$ for the EM with both $1$ hour and $1$ day resolutions. For the LCC we obtain $p=5$ (MSG), $p=2$ (EM5m), $p=2$ (EM10m), $p=3$ (EM30m), $p=1$ (EM1h and EM1d), respectively.
\item Contact networks display a very important link persistence pattern, as opposed to very small or zero cross interactions between links, at any time resolution, as verified by values of $q$ close to one, but $c$ close to zero (for both UCC and LCC), see the right panels of Fig.~\ref{figContact}. In fact, this social network is an example of the stability pattern characterising some social ties, such as friendship. Furthermore, such social system are described by Markovian dynamics ($p=1$).
\item Football networks display both link-specific persistence and cross interactions between links, with the two patterns which are inversely correlated as functions of the time resolution. For high resolution (1-2 sec), we measure large values of $q$, as opposed to small values of $c$. This is the result of contacts between players lasting for longer than the typical resolution and resulting in links persistent over several network snapshots, thus described by the mechanism of copying (itself) from the past. However, when time resolution becomes lower than the typical duration of a contact, link-specific persistence patterns disappear, in favour of some cross interactions between links, probably related to game strategies in football which appear evident at that specific time scale. In particular, this behaviour results in lagged cross correlations for the link dynamics which are thus captured by large values of the parameter $c$. This is an example of how including cross-interactions is crucial to capture the dynamics of the system. This is further confirmed with a simple exercise of link prediction, see below. 
Finally, when time resolution is too low (1 min), the temporal information is destroyed and the estimated parameters $q$ and $c$ are small or close to zero for almost all datasets. For the football networks, we do not notice much difference between the UCC and the LCC coupling models, because of an almost full backbone graph. Finally, the order of the CDARN(p) model is selected equal to $p=1$, for all football matches at any time resolution.
\end{enumerate}
}

\subsection{\piero{Heterogenous and time-varying patterns in real networks}}
\piero{Real-world networked systems may display heterogenous patterns in link dynamics and stationary properties. First of all, the probability for the appearance of a link is, in general, link-specific. For example, transports connecting different parts of a city are more or less frequent depending on people traffic, thus there are many buses crossing the main streets, while few buses connect the periphery. Second, auto- and cross-correlations of links may in general differ link by link. For example, in football contacts between midfielders tend to be persistent in time since the game is played largely in the midfield, while any contact between the forward and the defensive players is likely to be quick and short, both for scoring and defence.}

\begin{figure*}
\centering
\includegraphics[width=0.99\textwidth]{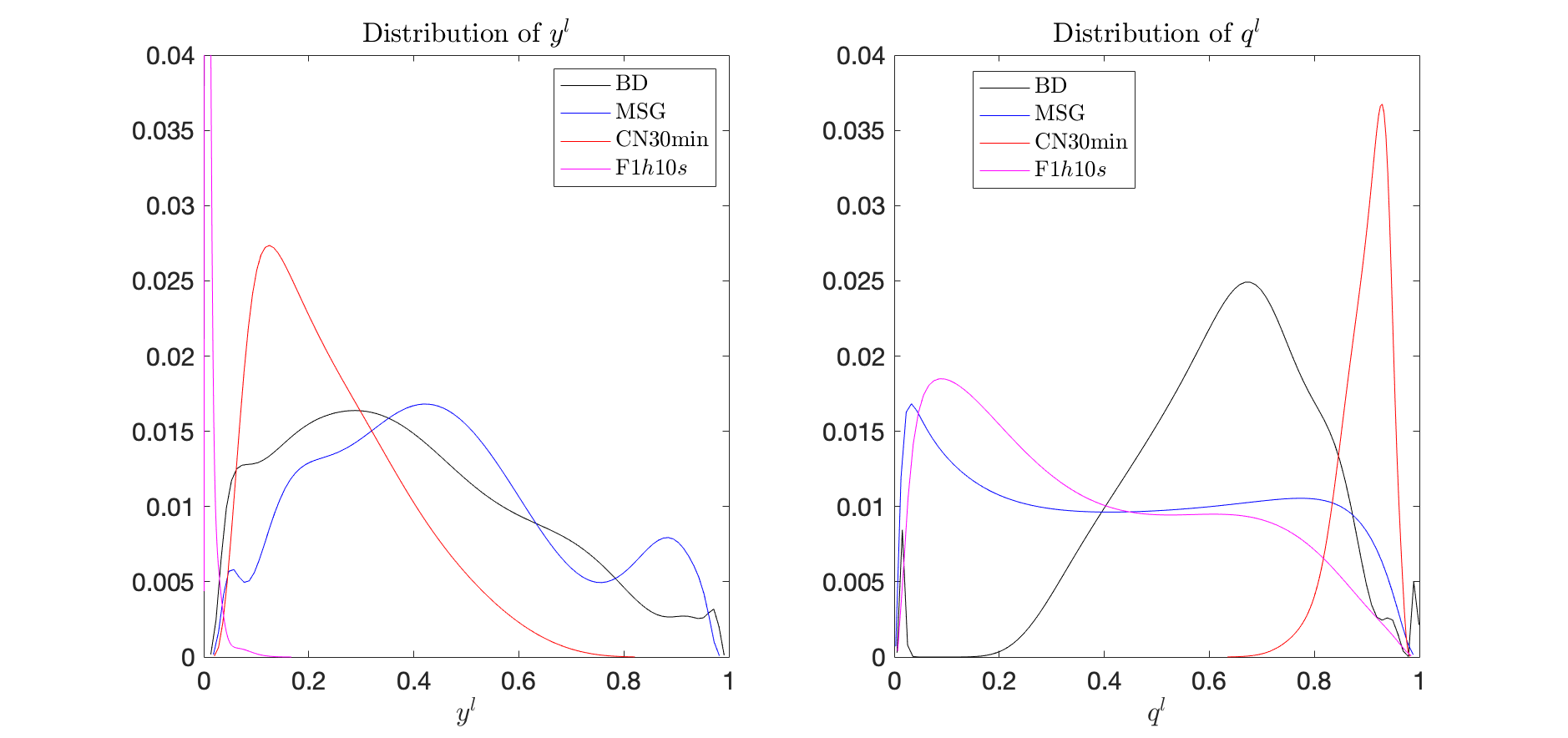}
\caption{\piero{{\bf Distribution of $y^\ell$ (left) and $q^\ell$ (right)} of the {\it heterogeneous} CDARN(1) (LCC) model estimated on temporal network data built for
    four datasets as indicated in the legend. Parameters are estimated by solving Eqs. (\ref{mle_cdarnp_eq_Hy}) and (\ref{mle_cdarnp_eq_Hq}), respectively.}}
\label{figHpatterns}
\end{figure*}

\piero{In our framework, we can study such behaviors in link dynamics
  by considering the version of CDARN model with {\it heterogenous}
  parameters. In order to point out the relevance of the heterogeneous
  generalization, in the following we consider the Local Cross
  Correlation coupling CDARN(1) model with Markovian memory (as
  suggested by previous results) with heterogeneous parameters $y^\ell$
  and $q^\ell$ applied to four network datasets: BD, MSG, CN30min, and
  F1h10s. The estimation method in such cases is described in the
  Appendix Section \ref{sm_hetero_mle}. The results are shown in
  Fig.~\ref{figHpatterns}. It is interesting to notice that some
  networked systems display a similar marginal link probability among
  links, such as football or contact networks, while others, such as
  transportation and online social communication networks, are
  characterized by some degree of heterogeneity. A similar result is
  obtained by looking at the autocorrelation structure of networks,
  with similar autocorrelations of links for transportation and
  contact networks, while link-specific autocorrelation properties are
  observed in the other cases. An analysis such that suggests time by
  time when the approximation with global parameters is enough or not
  for the precise description of a given network dataset.}

\begin{figure*}
\centering
\includegraphics[width=0.99\textwidth]{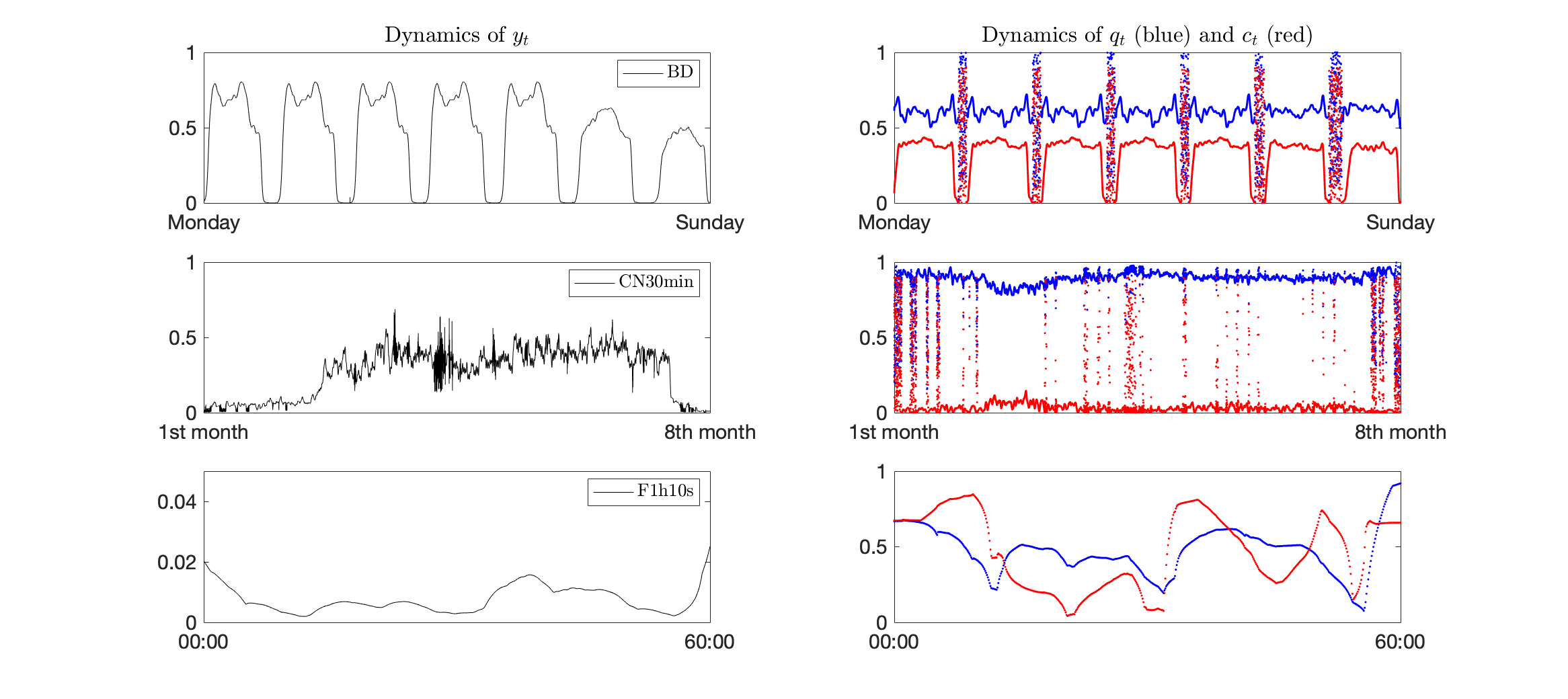}
\caption{\piero{{\bf Estimated dynamics of time-varying parameters of the CDARN(1) model}, \ie $y_t$ in the left panels, while $q_t$ and $c_t$ in the right panels, by using local likelihood methods, as explained in the main text, for $3$ network datasets: BD, CN30min, and F1h10s. }}
\label{figTVpatterns}
\end{figure*}

\piero{Heterogeneity in networked systems 
    can be {\it spatial} as
  well as {\it temporal}. In general, the correlation structure of a
  network as well as link probability may change over time, thus
  displaying time-varying patterns in link dynamics. This behavior can
  be captured by using the CDARN model with time-varying
  parameters. In particular, we consider the LCC coupling model with
  Markovian memory and exploit local likelihood methods to estimate
  the dynamics of parameters. The results for three network datasets
  (BD, CN30min, and F1h10s) are shown in Fig.~\ref{figTVpatterns}.
  Such method is able to capture the ({\it
    smooth}) dynamics of the marginal link probability, see left
  panels of Fig.~\ref{figTVpatterns}, similarly to the results of
  the previous section. Moreover, now we are able to describe the
  time-varying patterns of both auto- and cross-correlations of link
  dynamics, see right panels of Fig.~\ref{figTVpatterns}. It is
  interesting to notice that some systems, such as transportation and
  contact networks, display quite constant (around some mean value)
  correlation structure (except for the periods of no link activity
  when the estimation results as noisy, \eg during night hours for
  transports). On the contrary, systems like football networks display
  significant time-varying patterns of link correlations, likely
  related to the different phases of the game.}

\subsection{\cg{The role of cross-interactions \piero{and time-varying patterns} in link prediction}}

\cg{Cross-interactions of links shape the dynamics of
real-world networks in many ways. A preliminary indication of this comes from the
values of the key parameter $c$ we have obtained above.} \piero{Moreover, real-world systems may display non-stationary patterns in link density as well as link correlations, which can be captured by time-varying parameters, as shown above.}
\piero{Such effects are significant not only for description, but also for forecasting}. \cg{This can be made more evident by devising a simple study of link prediction
in empirical networks based on the CDARN(p) model with constant and homogenous parameters $q$, $c$, and $y$, \piero{opposed to the case of heterogenous or time-varying parameters,} as follows.}

\cg{Assume that we have observed a temporal network up to time $t$ (and
also that the backbone does not change in time) and to try to predict
the appearance of a link $(i,j)$ at time $t+1$ based on the
information up to time $t$. The one-step-ahead {\it forecast} (or {\it
  prediction}) is defined as the probability projected at time $t+1$
of observing the link $(i,j)$, that is
\begin{equation}\label{lp_cdarn_maintext}
S_{t+1}^{ij} \equiv \mathbb{P}(a_{t+1}^{ij}=1\vert \{\underline{\underline{A}}_s\}_{s=t,t-1,...,t-p+1},q,c,y),
\end{equation}
\piero{for the CDARN(p) model with homogeneous and constant parameters. In the case of heterogenous parameters, it is
\begin{equation}\label{lp_cdarn_maintext2}
S_{t+1}^{ij} \equiv \mathbb{P}(a_{t+1}^{ij}=1\vert \{\underline{\underline{A}}_s\}_{s=t,t-1,...,t-p+1},q^{(ij)},c,y^{(ij)}),
\end{equation}
with link-specific parameters, as described above. In the case of time-varying parameters, it is
\begin{equation}\label{lp_cdarn_maintext3}
S_{t+1}^{ij} \equiv \mathbb{P}(a_{t+1}^{ij}=1\vert \{\underline{\underline{A}}_s\}_{s=t,t-1,...,t-p+1},q^{t},c,y^{t}),
\end{equation}
making sure to use a causal kernel (\ie weighting only observations up to time $t$) in the estimation procedure.}
For the related explicit formulas, see the Appendix Section \ref{sm_forecast}.
The time series of forecasts $\{S_t^{ij}\}$, together with the
  realisations $\{a_t^{ij}\}$, allow us to characterise the forecasting
  performance of the model by using some binary classifier. Here, we
  consider the Receiving Operating Characteristic (ROC) curve
  \cite{hastie2009elements}, which is the plot of the
  True Positive Rate (TPR) ({\it sensitivity}) against the False
  Positive Rate (FPR) ({\it specificity}) at various threshold
  values. In practical terms, the better the model performs in the
  forecasting, the higher the associated ROC curve is in the unit
  square, or, equivalently, the larger the Area Under the Curve (AUC),
  see the Appendix Section \ref{sm_forecast} for further details.
\begin{figure*}
\centering
\includegraphics[width=0.99\textwidth]{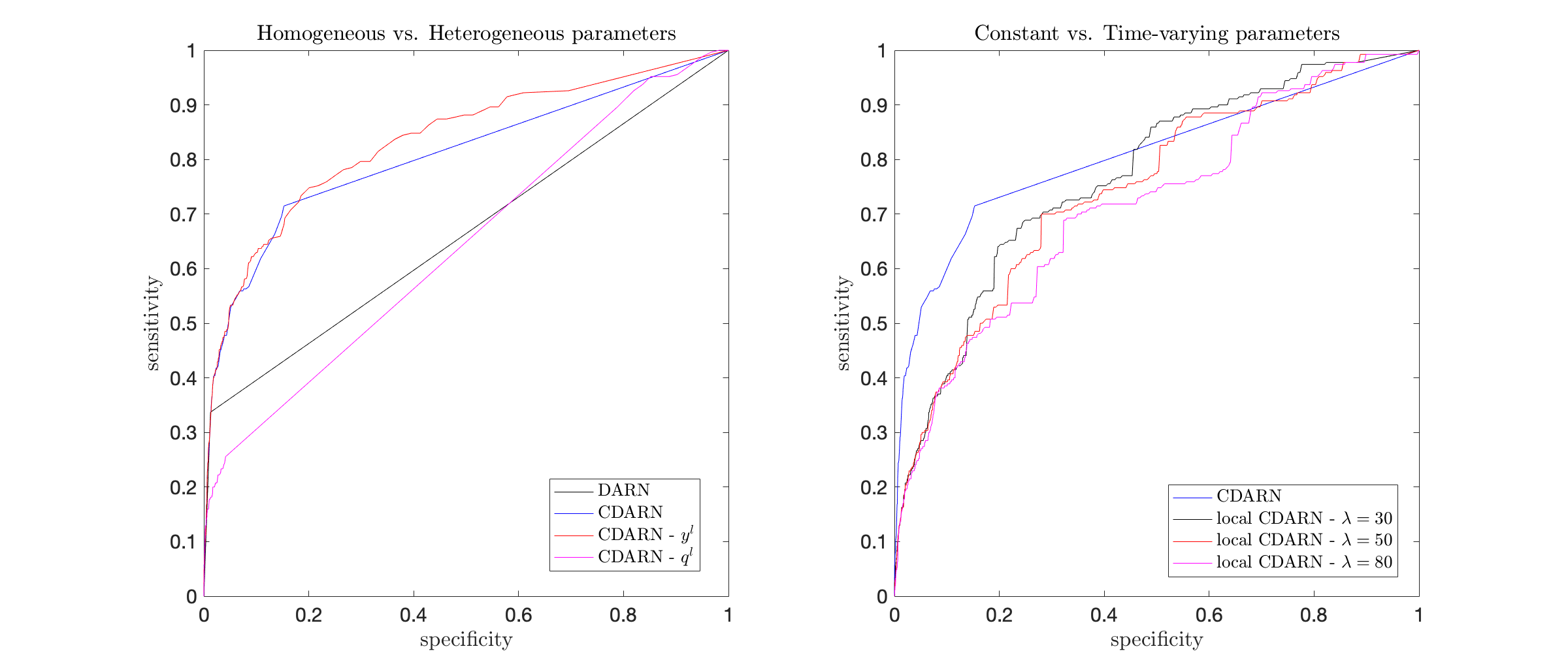}
\caption{\piero{{\bf Receiving Operating Characteristic curves} built (as described
  in the main text) for the DARN(1) model (LCC) and the CDARN(1) with NCC specification, applied to the football network associated with game 2 -
  away with time resolution of 10 sec. We compare the standard version of the CDARN model also with both the heterogeneous and the local generalizations, as described in the main text. In the case of time-varying parameters, we use the {\it causal} Epanechnikov quadratic kernel (\ie weighting only past observations) for different bandwidths $\lambda$.}}
\label{figRoc}
\end{figure*}
}

\cg{As case study we have considered the football matches network data set. 
%
We will show the results of the
prediction analysis for the network of game 2 away, although similar results
have been obtained for other matches.  
We aim to validate the model performance, in particular to verify the
effect of including cross interactions to better capture the network
dynamics of real-world systems, \piero{together with the role played by heterogenous and time-varying patterns in link prediction}. Thus, we compare the No Cross Correlation
(NCC) coupling model, \ie the DARN model, with the Local Cross Correlations (LCC) specification of the CDARN model, \piero{with either homogeneous, heterogeneous, or time-varying parameters}.}

\cg{The link prediction study is as follows: (i) we split the sample
  period in two, the first half of the match is used as a {\it
    training set} and the second half as {\it out-of-sample period},
  then (ii) we estimate the parameters of each coupling model on
  network data of the first half, by solving the MLE problem
  (\ref{mlecdarn}) for $q,c,y$ with data
  $\{\underline{\underline{A}}_t\}_{t=1,...,T^{half}}$ (\piero{or the
    corresponding problems for heterogeneous or time-varying
    parameters}), finally (iii) we construct the time series of
  forecasts, snapshot by snapshot, by considering a rolling window
  over the second half, \ie from $T^{half}+1$ to $2T^{half}$, thus
  obtaining $\{S^{\ell}_t\}^{\ell\in
    B}_{t=T^{half}+1,...,2T^{half}}$. Notice that in this exercise the
  model parameters $q,c,y$, \piero{or $q^\ell,c,y^\ell$ in the heterogenous
    case}, are estimated by using only data from the first period, and
  not updated each time the window rolls over new snapshots of the
  second half. \piero{On the contrary, in the case of time-varying
    parameters, every time the window rolls over a new observation the
    estimate of $q^t,c^t,y^t$ is updated}. The link prediction
  exercise is restricted to all pairs which can be connected on the
  backbone. In conclusion, we compare the time series of forecasts
  $\{S^{\ell}_t\}^{\ell\in B}_{t=T^{half}+1,...,2T^{half}}$ with the
  realisations $\{X^{\ell}_t\}^{\ell\in B}_{t=T^{half}+1,...,2T^{half}}$, by
  evaluating the ROC curve.  The results are summarised in
  Fig.~\ref{figRoc}, \piero{for match 2-away with time resolution
    equal to 10 sec (however, similar results are obtained for
    different matches and time resolution).  We can notice that the
    LCC coupling specification of the CDARN model (blue line in the
    left panel of Fig.~\ref{figRoc}), which accounts for both
    auto-correlations and cross interactions, always outperforms the
    DARN model (black line), accounting only for the auto-correlation
    of links. Moreover, accounting for the heterogeneity pattern of
    link probability $y^\ell$ (red line) plays an important role in link
    prediction of football data, largely outperforming the case with
    heterogeneous correlations $q^\ell$ (magenta line). The
    underperformance of the CDARN model with heterogeneous $q^\ell$
    parameters w.r.t. the DARN model is a signal of overfitting for
    the specific case of football networks. Finally, when comparing
    the forecasting performances of CDARN with constant or
    time-varying parameters, see the right panel of Fig.~\ref{figRoc}, the benefit of accounting for time-varying patterns
    in link prediction depends on the degree of specificity, \ie false
    positives, we are willing to accept to obtain some given degree of
    sensitivity, \ie true positives. In any case, a more timely
    estimation of parameters, associated with a tighter kernel
    bandwidth $\lambda$, tends to produce a better forecasting.}
In conclusion, the football network is an example
of the importance of taking into consideration lagged
cross-correlations of links, \piero{together with heterogenous or time-varying patterns}, in the description of the dynamics of
networked systems.
}

\section{Diffusion processes on correlated temporal networks}
\label{diff_process}

\cg{One of the most important points when modelling a networked system is
understanding how information, or some other quantity, spreads
throughout the system, in particular the rate of the diffusion and the
time in reaching the equilibrium. When links between nodes change over
time, then the first interest is on the role either memory and link
dynamics play in the diffusion process.}

In order to study this in a systematic way, in this Section we will
exploit the flexibility of the CDARN($p$) model introduced in Section
\ref{section2} which, as shown in Section \ref{empcdarn}, allows to
generate realistic temporal networks. The model allows to fine tune
the strength and length of the memory, while also controlling a key
feature of real-world networks, namely correlations between the
evolutions of links over time, as produced by dependencies between
their dynamics.

\subsection{Quantifying diffusion on a temporal network}

Diffusion is, in its original sense, the physical process by which
atoms and molecules move from regions of high concentration to regions
of low concentration. This process has been seen as an analogue to
processes in several other areas, such as opinion formation
\cite{watts2007influentials}, the motions and social interactions of
people \cite{schweitzer2007brownian}, and the movements of capital
through a financial system \cite{di2003exchanges}, and as such is
amongst the most common ways of describing spreading phenomena in
these areas. Indeed, diffusion finds uses in many other areas, where it 
is used as a linear approximation to non-linear systems, such at the 
Kuramoto model \cite{arenas2008rev}. 
\\
Complex networks often form the backbone of many real world
systems, and so it is natural to study diffusion over them 
\cite{boccaletti2006complex,newman2010networks}.  
In a diffusive process on a network the flow of information, or some
material, over a link is proportional to the difference in its
concentrations at the two nodes. The natural way to study
diffusion on a network is in terms of
the so called Laplacian matrix, which forms the network analogue of
the Laplace operator, which governs continuous time, continuous space,
diffusion.
%
Suppose we have a static undirected network with $N$ nodes and 
adjacency matrix $ \underline{\underline{A}}= \{ a^{ij} \}$.
The equation that governs the diffusion of some node related quantity  
$\underline{d}(s) \in {\mathbb R}^N$ over (continuous) time $s$
can be written as: 
\begin{eqnarray}
	{\underline {\dot{d}}}(s) = -\mu  \underline{\underline{\mathcal{L}}} \underline{d}(s) 
	\label{eqn:static_diff}
\end{eqnarray}
where $\mu$ is the diffusion coefficient, which controls the time
scale of the diffusion process, and  $\underline{\underline{\mathcal{L}}} = \{\mathcal{L}^{ij}\}$
is the graph Laplacian matrix, whose entries can be written in terms
of the entries of $ \underline{\underline{A}}$ as $\mathcal{L}^{ij} = \delta(i,j) k_i - a^{ij}$,
where $k_i =\sum_j a^{ij}$ is the degree of node $i$
\cite{newman2010networks}. Notice that this equation is in continuous
time; as a convention when a variable is continuously dependent on
time $s$, the time will be in brackets (e.g. $ \underline{\underline{A}}(s)$), and for discrete
time $t$ it will be given as an index (e.g. $ \underline{\underline{A}}_t$).  On a temporal network
the only thing that needs to be changed in this equation is that the
Laplacian matrix must be allowed to vary over time, hence $ \underline{\underline{\mathcal{L}}}
\mapsto  \underline{\underline{\mathcal{L}}}(s)$ where $ \underline{\underline{\mathcal{L}}}(s)$ is the Laplacian matrix
associated with the continuous time adjacency matrix $ \underline{\underline{A}}(s)$. 
This system exists in 
continuous time, and so the temporal network that underlies it
must also exist in continuous time.
The solution of the above equation is then clearly
$$
\underline{d}(T)=\exp\left(-\mu \int_0^T  \underline{\underline{\mathcal{L}}}(s) ds\right) \underline{d}(0)
$$

However,
the vast majority of models for temporal networks are discrete in time, and so, 
 given a model for a discrete time
temporal network, we must first embed the network in continuous time.
To this end, we assume that the adjacency matrix changes at discrete time steps of length $\Delta t$, taken, without loss of generality, to be equal to $1$. Thus the Laplacian $ \underline{\underline{\mathcal{L}}}(s)$ is piecewise constant and, according to the above notation, will be denoted by $ \underline{\underline{\mathcal{L}}}_t$ ($t=1,..,,T)$. The solution of the diffusion equation hence becomes 
\begin{equation}
	\underline{d}_{T} = \exp \left( -\mu \sum_{t=1}^{T}  \underline{\underline{\mathcal{L}}}_t \right) \underline{d}_0.
\end{equation}

As stated, our purpose here is to study the effects that memory in a
temporal network has on diffusion over that network. This is a very
general aim, and so we must be more specific about what we wish to
analyse.  Rather than studying spreading in terms of the full dynamics
of diffusion on a temporal network, i.e. the concentrations
$\underline{d}_t$ of material at each node at each time step $t$, we
can instead ask about how long it takes for this diffusion to reach
equilibrium.
In particular, since the changes in the network are responsible for any
changes in the rate of spreading, we focus on the number of 
network evolutions (number of time steps $\Delta t$) before equilibrium. 
To formalise this concept we first note that in general
we will not reach equilibrium in a finite number of timesteps, and so we instead fix some small
 positive $\epsilon$, so that the {\em time to equilibrium} is then defined as:  
\begin{equation}
	\tau = \min_{t \in \mathbb{N}} (t : \left| \underline{d}(t) - \underline{u} \right| < \epsilon), 
\label{eqn:eqm_condition}
\end{equation}
where the vector $\underline{u}$ is the uniform vector with $u^i = 1/N$, 
which corresponds to the equilibrium state of the diffusion process 
on a connected network with $N$ nodes.
For our purposes the norm $| \cdot |$ will be taken to be the
Euclidian norm.
For our purposes we will keep the value of $\epsilon$ fixed as
$\epsilon = 10^{-3}$. 
The temporal networks we will use here are generated
by discrete-time random processes, and so $\tau$ will be a random
variable.  Given this, we will focus on finding the average of this
value, $\left<\tau \right>$, over several realisations of the
system. Unfortunately, $\left<\tau \right>$ will be highly dependent
on the structure or size of any temporal network being studied, and so
it would be impossible to draw conclusions about the influence of any
model parameters in these systems. Our goal here is to study the
effects of memory on spreading rate, and so we must introduce some way
of comparing the time to equilibrium as a function of this memory as
given by different networks.  To this end we normalise $\tau$ by
expressing it in terms of the time taken for a diffusion to reach
equilibrium on the same backbone, but with a memoryless temporal
network. In other words, we define the {\em rescaled time to equilibrium}, 
$\mathfrak{T}^p$, given memory length $p$, in terms of $\left<\tau^p \right>$, 
the average time to equilibrium given memory length $p$, as:
\begin{equation}
	\mathfrak{T}^p = \frac{\left< \tau^{p} \right>}{\left<
          \tau^{0} \right>} .
	\label{re_scaled_eqm_time}
\end{equation}
 Notice that the CDARN($p$) model does not directly allow for $p=0$, and so we 
 define $\tau^{0}$ to be the case where $q=0$, and so no memory is ever used.
 This allows us to compare the effects that changing the memory length $p$ and
 the coupling matrix $ \underline{\underline{C}}$ have on different backbones.

\subsection{Numerical results}
\label{sec:nr}

We have first investigated the rescaled time to equilibrium of
  a diffusion process on CDARN($p$) temporal network models with different
  backbones by means of an extensive set of numerical simulations.
The value of the parameter $\mu$ allows to tune the time scale of the
diffusion process, while the three parameters controlling the
link density $y$, memory strength $q$, memory length
$p$, and the two matrices network backbone $ \underline{\underline{B}}$, and link coupling
matrix $ \underline{\underline{C}}$, control the properties of the temporal network.
To construct the backbones $ \underline{\underline{B}}$ we have taken 
three real-world temporal networks, each with different structural
properties, and we have aggregated their links over the 
extent of the available network data and discarded  
the link weights. The three real temporal networks we have 
considered are: 
(i) Flights between US airports (Airport) \cite{usairport_bb}. 
(ii) Email interactions between employees at a manufacturing company (Email) \cite{email_bb}.
(iii) Journeys on the London underground (Tube) \cite{ldn_tube_bb}.
The key features of the three resulting 
backbones are summarised in Table \ref{tab:bb_features}.
The number of nodes in the three networks ranges from about 100
  to 300. With 302 nodes and an average degree $\left< k \right>=2.3$
  the Tube is the backbone with the smallest link density, while Email
  is a very dense backbone with links connecting $23 \%$ of the possible
  pairs of nodes.
\begin{table}
\centering
\begin{tabular}{|c|c|c|c|c|c|}
\hline
{\bf Backbone}& {\bf $N$} &{\bf $\left< k \right>$} &{\bf $D$} &{$\lambda_N$} &{$\lambda_2$}\\
\hline
Airport& 143 & 2.030 & 0.0143 & 31.02 & 0.01696\\ 
\hline
Email& 167 & 38.93 & 0.2345 & 140.0 & 0.3811 \\
\hline
Tube& 302 & 2.311 & 0.0078 & 8.432 & 0.005918 \\
\hline
\end{tabular}
\caption{ \label{tab:bb_features} {\bf Key structural features for
    each backbone.} The number of nodes ($N$), average degree ($\left<
  k \right>$), density ($D$), and dominant ($\lambda_N$) and smallest
  non-zero (spectral gap, $\lambda_2$) eigenvalues of the Laplacian matrix for each
  backbone. }
\end{table}
Our aim here it to study not only the effects of memory, but also the interplay 
between memory and correlations in the dynamics of links.
In order for us to clearly observe the effects of these features we 
must be able to compare different models: one in which the evolution 
of links is correlated, and one in which links are independent.
As such we have simulated our system on each of the three
different backbones $ \underline{\underline{B}}$ for a range of different parameters $p,q,y$ and $\mu$,
and, for the three different forms of the coupling matrix $ \underline{\underline{C}}$, see Section \ref{cdarn_description}.

\begin{figure*}
	\includegraphics[width=1.0\textwidth]{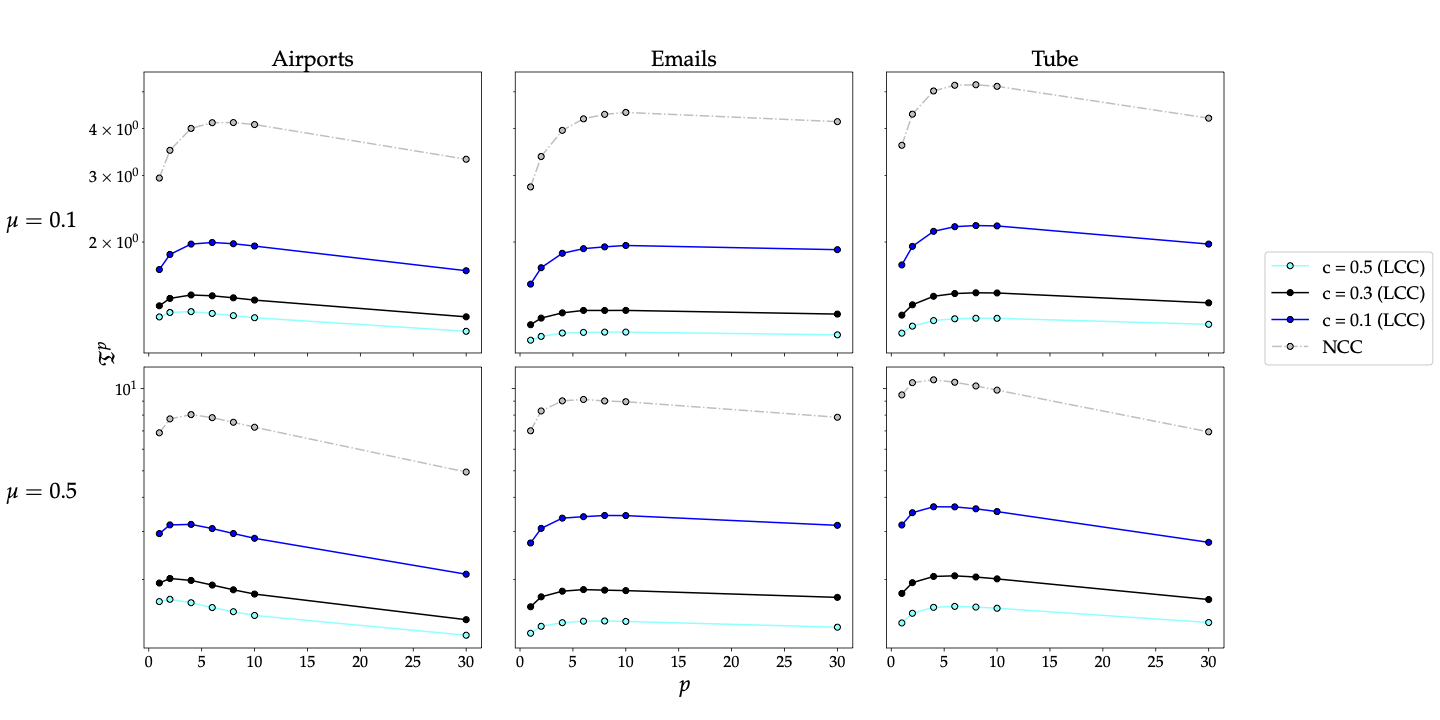}
	
	\caption{{\bf Rescaled time to equilibrium for diffusion on different network backbones as a function of the memory
          length} $p$ for a CDARN(p) model with local (solid lines, LCC), and no (grey line, NCC) cross correlations between different links.
          Memory strength $q$ is kept constant at 0.95 to ensure that memory plays a significant role in the evolution of the network and
          link density $y$ is kept at 0.1 to ensure that there is sufficient time for any effects of memory to be observed.
          The coupling strength $c$ and diffusion speed $\mu$ are varied. The backbones were taken from a collection of real data sets. Averages were taken over $2\cdot 10^4$ realisations of the process. Note that a semi-log scale has been used.}
\label{fig:bb_speedup}
\end{figure*}

In our simulations, for each instance of diffusion on a CDARN($p$) model,
i.e. for each different set of parameters $\mu$, and 
$p,q,y, \underline{\underline{B}}, \underline{\underline{C}}$ and $c$, we compute $\mathfrak{T}^{p}$. 
This is done directly by estimating 
$\left<\tau^{p} \right>$ and $\left<\tau^{0} \right>$, where the averages are
taken from multiple realisations of the diffusion process. 
 In each case the initial condition for the diffusion $\underline{d}_0$ is 
 such that all of the material to be diffused is placed at a random node $j$: 
 $d_{0}^{i} = \delta(i,j)$ where $j \in \{1,...,N\}$. In this way we avoid 
 any bias that might be introduced by repeatedly choosing the same 
 starting node. Before any diffusion takes place on the temporal network,
 we allow the CDARN($p$) temporal model to evolve until it has reached a steady state
 (see Appendix G). 
 
In Fig.~\ref{fig:bb_speedup} we report the rescaled time to
equilibrium $\mathfrak{T}^{p}$ given memory length $p$ as a function
of $p$, for each backbone and with a number of different sets of model
parameters. Note that a semi-log scale has been used. 
To ensure that memory plays a significant part in the
evolution of the temporal network we have fixed the memory strength $q =
0.95$, and to ensure that the system has enough time for the effects
of memory to be observable we have fixed the link density $y=0.1$.  We then
vary the diffusion speed $\mu = 0.1, 0.5$.
All of these results are shown for both the local cross correlation model,
with the three values of the coupling strength
$c = 0.5, 0.3, 0.1$, and the no cross correlation model, i.e. the
case $c=0$. 
For $\mu=0.1$, and hence slow diffusion, we observe that the equilibrium
time is non-monotonically dependent on the memory length for all
backbones, coupling strengths, and for both coupling models. This
non-monotonicity is most prominent when $c=0.1$, but far less so when
the coupling is stronger. When we consider $\mu = 0.5$, and hence faster
diffusion, the observed non-monotonicity is far less apparent in all
but the NCC model, there is however still a  clear
dependence on
memory, particularly for lower values of $c$.  Unsurprisingly, there
is a significant difference between the results for the no correlation
model and those with correlations: in all cases local correlations
speed up diffusion. What we do notice though is that there is no
marked difference between different backbones. Since we have
normalised each set of results this is not entirely unexpected.
 

In summary, the rescaled equilibrium time shows a number of interesting 
features as a function of the memory length $p$, the coupling 
matrix $ \underline{\underline{C}}$ and the backbone $ \underline{\underline{B}}$. Most notable among these features 
are :
\begin{itemize}
\item The rescaled time to equilibrium $\mathfrak{T}^{p}$
  is generally a non-monotonic function of the memory length $p$.

\item Stronger local correlations, i.e. larger values of the
  coupling strength $c$ speed up diffusion. 

\item Correlations have a considerable effect on the influence of memory:
  when the coupling strength $c$ is high then diffusion properties are weakly dependent on the memory properties of the network.

\end{itemize}

As we will show in the following,   
by understanding the behaviour in the limit of no cross correlations, and by isolating the effects 
of temporal correlations, we can get a clear picture of the causes of
our observations.

\subsection{Analytical results in the no cross correlations limit}
\label{sec:trA}

In light of our numerical results, we now study the 
theory which underpins both the CDARN($p$) model
and the diffusion of material over it.
We first study diffusion on the simplest form of the CDARN($p$) model,
the limit of no cross correlation between the dynamics of links.
This is precisely the NCC coupling model
that was previously introduced. In such a limit
the links of the CDARN($p$) model are independent
processes, and so we can study them in isolation. 
In this case, as we will show below, the model is
  analytically tractable and it is possible 
to derive
an analytical expression for the rescaled time to
equilibrium.

In order to analyse the dynamics of diffusion over a single link of
the CDARN($p$) model, let us consider two nodes, one of which has an
amount of a material, and the other of which has some other amount.
The diffusion of this material out of the first node is given by: 
\begin{eqnarray}
    \dot{d}^1(s) = -\mu \left( d^1(s) -d^2(s) \right) a_t^{1\,2}.
    \label{eqn:two_node_diff}
\end{eqnarray}
where $t = \floor{s}$, and the random variable $a_t^{1\,2}$
describes the presence of the link between node 1 and node 2
at discrete time $t=0,1,\ldots$ as governed by the DAR($p$) process
defined in Eq.~\ref{DARp_eqn}. 
When combined with the conservation condition $d^2(s) = 1 - d^1(s)$ 
this describes the full dynamics of the diffusion process.
Given any set of initial conditions we can first find
the number $\tau$ of time steps before equilibrium is reached.
By noticing that, since when the link is not present there can be no diffusion, we only 
need to count the number of times that the link is present.
If we were to take 
$a_t^{1\,2} = 1$ for all $t$, then we can easily find $\tau = n$, 
and express $n$ in terms of 
$\mu, \Delta t$ and $\epsilon$ (see Appendix D). 
 Now let us associate with 
$a_t^{1\,2}$ the counting process $F_t = \sum_{k=0}^t a_k^{1\,2}$. 
We can then see that for a link that changes in time  
$\tau = \min_{t>0} (t : F_t = n)$.  This allows us to re-phrase our 
problem: we now want to find the average time taken until a link governed 
by a DAR($p$) process has occurred $n$ times. 
The DAR($p$) process that governs the link can be thought of as a 
$p$-th order Markov process with the following transition matrix
(see Appendix E for a full explanation and discussion):
\begin{eqnarray} \label{transition_mat}
	T_{\alpha \beta} = \left[q \frac{h(\alpha)}{p} +(1-q)y \right] \delta \left(\beta,2^{p-1} + \floor{\frac{\alpha}{2}}\right) +\nonumber \\
	 \left[1 - q \frac{h(\alpha)}{p} - (1-q)y \right] \delta \left(\beta,\floor{\frac{\alpha}{2}}\right) 
\end{eqnarray}
Here $\alpha$ and $\beta$ represent some indexing of the $S = 2^p$ possible memory states,
 $h(x)$ is the Hamming weight of the number $x$ (the number of 1's
in its binary representation), $\delta(x,y) = 1$ if $x = y$ and $0$
otherwise, and $\floor{x}$ is the largest integer value smaller than
$x$.
If we break this matrix up into two parts, $ \underline{\underline{T}}^L =  \left[1 - q \frac{h(\alpha)}{p} - (1-q)y \right] \delta \left(\beta,\floor{\frac{\alpha}{2}}\right)$
and $ \underline{\underline{T}}^R =  \underline{\underline{T}} -  \underline{\underline{T}}^L$, then we can find the average time $k_{\alpha} \in \mathbb{R}_{\geq 1}^S$ 
taken for a link to occur 
given that it started in state $\alpha$ as \cite{cinlar2013introduction,ballester2014random}
\begin{equation}
	\underline{k} = \left(  \underline{\underline{I_d}} - \underline{\underline{T}}^L \right)^{-1}  \underline{1},
	\label{avg_hit_time}
\end{equation}
and the probability $h_{\alpha \beta}$ that when a link occurs it will occur in state $\beta$, given 
that it started in state $\alpha$ as
\begin{equation}
	\underline{\underline{h}} =  \left(  \underline{\underline{I_d}} - \underline{\underline{T}}^L \right)^{-1}  \underline{\underline{T}}^R .
\end{equation}
Now let us define $\omega_{\alpha} $ as the probability that a link starts in state $\alpha$. 
We can then find the average time taken until the $n-$th link in a $p-$th order system as:
\begin{equation}
	\left< \tau^p \right> = \underline{\omega}^T \left( \sum_{t=0}^{n-1} \underline{\underline{h}}^t \right) \underline{k}.
\end{equation}
%
We now have an explicit formula for the average number of time steps
to equilibrium. 
However, it is impossible to compare values of $\left< \tau^p \right>
$ directly, as such values will be heavily dependant on parameters of
the model other than the memory length $p$.
Because of this, we look at the rescaled time to
equilibrium as defined in Eq.~\ref{re_scaled_eqm_time}.
The limiting behaviour of this quantity can be studied analytically.
 First, we note that $\left< \tau^0 \right>$ can be found 
directly as $n/y$. Then,
we observe that as $p \to \infty$, $\left< \tau^p \right> \to n/y$ (see 
Appendix C and H), meaning that our large
memory limit is exactly the same as the no memory case, and 
because of this $\mathfrak{T}^p$ is not intrinsically bounded above (see appendix H).
We can also directly solve for $p=1$, and in principle extend these calculations to solve for small $p$ (see appendix H).
 Finally we can show that, when $y$ is ``small enough", as it is in all of our cases, 
$\left< \tau^p \right> \geq \left< \tau^{\infty} \right>$, and hence that
\begin{equation} 
	\mathfrak{T}^p \geq 1.
\end{equation}
Hence the rescaled equilibrium time in the large memory limit acts as a
lower bound for the case of arbitrary $p$ (see Appendix I), explaining the
similar behaviour observed in the full system.  It should be noted
that in cases where $y$ is not ``small enough" we will observe the
opposite effect: the large memory limit will be an upper bound.

\begin{figure}
	\includegraphics[width=0.5\textwidth]{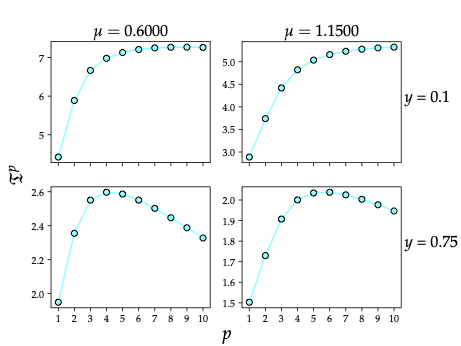}
	
	\caption{{\bf Rescaled time to equilibrium for diffusion over a link} 
          in the limit of the CDARN($p$) model with no cross correlations as a function of the memory length $p$. The dynamics of the link is
          generated by a DAR($p$) model with $q=0.95$, for various values
          of $y$. Two different values of diffusion constant $\mu$ were used.}
\label{fig:link_speedup}
\end{figure}

When plotting this rescaled time to equilibrium as a function of $p$ for
various $\mu$ and $y$, as in Fig.~\ref{fig:link_speedup}, we observe
many of the same traits we found in Section \ref{sec:nr} 
for the full CDARN($p$) model with cross
correlations. 
Principally, the following two similarities needs to be noted.
Firstly we see evidence for the previously explained large memory limit,
i.e. the rescaled time to equilibrium is bounded below by the value 
obtained in the limit of large $p$. 
Secondly we see that $\mathfrak{T}^p$ can be highly non-monotonic as a function of $p$. 

In summary, the study of the CDARN($p$) model in the limit of no-cross
  correlations provides us with a good understanding of the causes for 
two of the most notable phenomena observed in the full network systems, 
and allows us to focus on the role of correlations in inducing the
remaining effects. 

\subsection{Derivation of the temporal correlation matrix}
\label{sec:trB}

We will now present a general analytical approach to finding
the lagged cross and autocorrelations for an arbitrary coupling 
matrix $ \underline{\underline{C}}$, which we will use in Section \ref{sec:qec} to explore the interplay between 
correlations and memory in more depth than would be possible through simulations alone.
In particular, we will use it to isolate the effects that correlations among neighbouring links 
in the LCC coupling model have on the time taken for diffusion processes 
on the temporal network to reach equilibrium. 

The results of Section \ref{sec:nr} clearly indicate
that the presence of coupling in the temporal dynamics of different
links plays an important role in the behaviour of the rescaled time to
equilibrium for a diffusion process on a temporal network. 
Indeed our claim is that, while the temporal autocorrelations of links 
slow down diffusion 
\cite{masuda2013temporal,Scholtes_natcomm14,Rosvall_natcomm14,Hiraoka:2018aa,jo2015correlated},
as evidenced by the limit of no cross correlations case,
temporal correlations among neighbouring links speeds it up.
Fortunately, the CDARN($p$) model is analytically 
tractable enough for us to fully describe the correlations 
that are present for a general coupling matrix $ \underline{\underline{C}}$, without 
relying wholly on simulations. 

Rather than working with the backbone network directly, we will 
instead consider the corresponding graph in which 
links are the  nodes in the backbone, and we connect any two nodes in the new graph
if their links in the backbone graph shared a node. For a backbone network with $L$ possible 
links, we assign each of these links with a linear index.
Then let us denote the correlations 
between link $\ell$ and $\ell'$ at time lag $k$ as $\left< a^{\ell}_t a^{\ell'}_{t-k} \right> = \rho^{\ell \ell'}_{k}$.
Given a coupling matrix $ \underline{\underline{C}}$, $\rho^{\ell \ell'}_k$ can be found as 
the solution to the following Yule-Walker equations \cite{jacobs1978discrete,macdonald1997hidden}:
\begin{equation}
	\underline{\underline{\rho}}_k = \frac{q}{p}  \underline{\underline{C}} \sum_{a=1}^{p}  \underline{\underline{\rho}}_{k-a}.
\end{equation}
Note that we have dropped our indices, and so 
each element in the equation is a matrix.
We can show that, for general $ \underline{\underline{C}}$, this equation is
solved by the composition of  different functions over 
supports $k \in \{np,...,(n+1)p \}$ for integer $n$.
The first of these can be found to be constant, while 
the following are exponentially decaying (see Appendix J, K). Because of this,
we can characterise the correlations at all values of $k$ 
in terms of this initial constant, which we call $ \underline{\underline{\rho}}$.
We first define the following tensor:
\begin{equation}
	\Delta^{\ell \ell' \ell''} = \frac{q}{p} \left( (p-1) c^{\ell \ell''} + q \sum_{b \neq \ell'} c^{\ell b} c^{b \ell''} \right),
	\label{eqn:delta_val}
\end{equation}
then $\rho^{\ell \ell'}$ can be found as the solution to the following system of 
linear equations (see Appendix J) 
\begin{equation}
	\rho^{\ell \ell'} = \sum_{\ell''=1}^{L} \Delta^{\ell \ell' \ell''} \rho^{\ell'' \ell'} +  \frac{q}{p} c^{\ell \ell'}.
	\label{eqn:rho_gen_main}
\end{equation}
The system can be greatly simplified in special cases (see Appendix K, L, M, N).
For example, in the case of the UCC coupling model, we show that 
$\rho^{\ell \ell}$ is constant for all $\ell$, and similarly $\rho^{\ell \ell'}$ is constant 
for all pairs $\ell, \ell'$ such that $\ell \neq \ell'$, thus reducing the calculation of the 
correlation coefficients to solving a pair of linear simultaneous equations.
Given this set of equations for $\rho^{\ell \ell'}$, we can also then find the correlations $\left<a^{\ell}_t a^{\ell'}_t \right> = \rho^{\ell \ell'}_0$
when $\ell \neq \ell'$ as
\begin{equation}
	\rho^{\ell \ell'}_0 = q \sum_{\ell''=1}^{L} c^{\ell \ell''} \rho^{\ell'' \ell'}.
\end{equation}
This gives us a full picture of the correlations present in
the CDARN($p$) model and 
allows us to calculate them directly.

\subsection{Cross-interactions speed up the diffusion}\label{sec:qec}

We saw in our study of diffusion in the limit of no cross correlations
that the rescaled time to equilibrium $\mathfrak{T}^{p}$ is a non-monotonic function of
the memory length $p$.  It is also widely understood 
that a way of characterising memory of a time series is by using the autocorrelation function.
We find 
for a CDARN($p$) process the memory $p$ is precisely the value
for the time lag $k$ after which the correlation function 
$ \underline{\underline{\rho}}_k$ decays
exponentially (see Appendix H and results in \cite{Williams_2019}).
 With this in mind we can now focus on the comparison
between autocorrelation coefficients of links in a CDARN($p$) temporal
model and the cross correlation coefficients of neighbouring links.
This is done by studying the constant values for the auto and cross correlation coefficients
at time lags $k \le p$. 
In order to do this effectively for large networks we will average these
quantities over all links (and neighbours where appropriate) to gain
the averaged autocorrelation coefficient $\rho_{ac}$ and the averaged
neighbourhood correlation coefficient $\rho_{ncc}$. These are
defined, given the matrix of correlation coefficients $\rho^{\ell
  \ell'}$ for a backbone $ \underline{\underline{B}}$ with $L$ links derived in
  Section \ref{sec:trB}, as
\begin{eqnarray}
\begin{aligned}
	\rho_{ac} =& \frac{1}{L} \sum_{\ell=1}^{L} \rho^{\ell \ell},\\
	\rho_{ncc} =& \frac{1}{L} \sum_{\ell=1}^{L} \frac{1}{\left| \partial_B \ell \right|} \sum_{\ell' \in \partial_B \ell} \rho^{\ell \ell'}.
\end{aligned}
\label{eqn:rho_ac_ncc}
\end{eqnarray}
where as before $\partial_B \ell$ is the set of links in the neighbourhood of $\ell$
on the network backbone $ \underline{\underline{B}}$.\\

\begin{figure*}
	\includegraphics[width=1.0\textwidth]{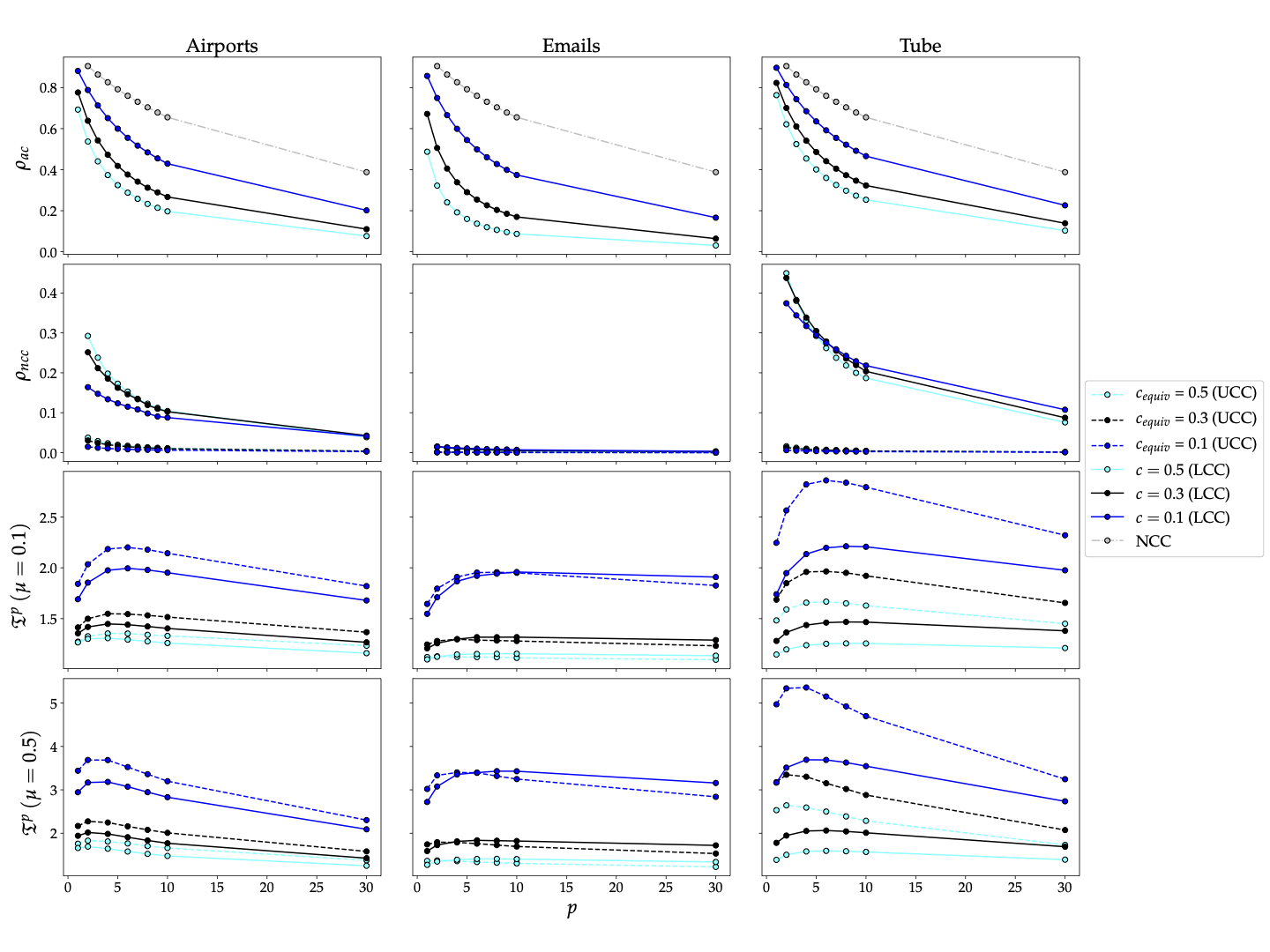}
	
	\caption{{\bf Average autocorrelation and neighbourhood correlation of a link, and rescaled time to equilibrium for diffusion} for both the LCC and UCC coupling models on different backbones as a function of the memory length $p$. The value of the average $\rho_{ac}$ (first row), and $\rho_{ncc}$ (second row), where averages are taken over links in a CDARN($p$) temporal network with local cross correlation (solid line, LCC),  uniform cross correlation (dashed line, UCC), and no cross correlation (dash/dot line, NCC) coupling, for each backbone. The third and fourth rows display the rescaled average time till equilibrium for a diffusion process on these networks with diffusion constants $\mu = 0.1$ and $\mu = 0.5$ respectively.
Note that UCC is not included in the first row ($\rho_{ac}$) as its values are, by construction, precisely the same as those of the LCC model. The NCC model is not included in the second row ($\rho_{ncc}$) as its value is always zero. 
Memory strength $q$ is kept constant at 0.95 to ensure that memory plays a significant role in the evolution of the network and link density $y$ is kept at 0.1 to ensure that there is sufficient time for any effects of memory to be observed. The coupling strength $c$ and diffusion speed $\mu$ are varied. Note that for the UCC model we assign the curves the value $c_{equiv}$ rather than $c$, this is because we chose the values of $c$ to match the value of $\rho_{ac}$ for the UCC and LCC coupling models, as such $c_{equiv}$ refers to the value of $c$ in the LCC model that is being matched. The backbones were taken from a collection of real data sets. The rescaled times to equilibria were averaged over $2\cdot 10^4$ realizations of the process.}
\label{fig:bb_corr_comp}
\end{figure*}

For clarity, let us now re-state our claim, as based on our observations of the numerical simulations
displayed in Fig.~\ref{fig:bb_speedup}:  while autocorrelation
of links slows down diffusion, correlations between neighbouring links
speeds up diffusion. 
While in Fig.~\ref{fig:bb_speedup} we do see that diffusion is 
faster in the LCC model than in the NCC model, 
the autocorrelations of links in the two models are different. Further to this,
while it would be possible to tune the parameters of the NCC model so that 
it produced links with the same autocorrelation coefficient as the LCC model, 
as the NCC model does not have a coupling strength, this could only be
achieved by changing either the memory strength $q$ or the memory length $p$.
Because of this we can not
judge the influence of neighbourhood correlations from our previous results,
 and we cannot use the NCC model to explore the effects of neighbourhood correlations further.
In order to give a valid point of comparison, we can now make use of our 
third coupling model, which allows us to precisely control the average 
link autocorrelation, but also removes any correlations between neighbouring 
links. 
To recall, for a backbone with $L$ links, the coupling matrix in the
UCC model $ \underline{\underline{C}} = \{c^{\ell \ell'}\}$ is given by $c^{\ell \ell'} = (1-c)
\delta(\ell, \ell') + (1-\delta(\ell, \ell')) {c}/({L-1})$.  The
simplicity of this model lends itself well to analytical calculations,
and so we can now use this model to isolate the effects of
neighbourhood correlations. Indeed we can show that in the limit
  of large numbers of links $L$ this model reduces to a DARN($p$)
  temporal network on a fixed backbone, in which links are independent
  (see Appendix N).  First we fix the parameters $p,q$ and $y$ for both
the LCC and UCC models, this ensures that there is the same
memory strength and length, and the average degree of the temporal
networks produced are the same. We can then fix the value of $c$
for the LCC model, as shown in Fig.~\ref{fig:bb_corr_comp} (first
row), and calculate the resulting value of $\rho_{ac}$ and
$\rho_{ncc}$, as defined by Eq.~\ref{eqn:delta_val},\ref{eqn:rho_gen_main} and \ref{eqn:rho_ac_ncc}.
By then varying the value of $c$ used in the
corresponding UCC model we obtain precisely the same value for
$\rho_{ac}$, while leaving $\rho_{ncc} \approx 0$, because in the considered network backbones the number of links $L$ is large.
In figure
\ref{fig:bb_corr_comp} we plot both the values for $\rho_{ac}$ and
$\rho_{ncc}$, along with the rescaled time till equilibrium for a
diffusion process on the corresponding temporal network.  Note that
the LCC and UCC models have, by construction, exactly the same value
of $\rho_{ac}$, and so only LCC is plotted in the upper panels, and
the NCC model must always have $\rho_{ncc} = 0$, and so it is not
plotted in the lower panels.

We observe here that, as expected, the value of $\rho_{ac}$ in the
NCC model is significantly higher than for the LCC model. We also see
that both $\rho_{ac}$ and $\rho_{ncc}$ decay as the memory length 
$p$ increases, consistent with the DARN($p$) temporal network model.
Most notably, there are significant differences between the values of 
$\rho_{ncc}$ given for each backbone. While both  the Airport and Tube 
backbones display significant neighbourhood correlations in the LCC 
coupling model (though the values are larger for the Tube backbone),
the Emails backbone
only has notable neighbourhood correlations for low values of $p$, and indeed 
at $p=30$ is practically indistinguishable from the NCC model. Also, as expected the 
value of $\rho_{ncc}$ for the UCC model is always approximately 0.
We have tested our hypothesis by comparing the rescaled time to equilibrium (see the lower 
two rows of Fig.~\ref{fig:bb_corr_comp}) 
$\mathfrak{T}^p$ for both the LCC and UCC models, both generated and plotted in
exactly the same way as was done for Fig.~\ref{fig:bb_speedup}.
%
It is clear that when the value of $\rho_{ncc}$ is large, as in the Tube backbone, 
diffusion on the LCC coupling model is always faster. When $\rho_{ncc}$ is 
lower, as in the Airports backbone, this is still true. Finally, when there are little 
to no correlations between neighbours, as with the Email backbone, diffusion on
the LCC coupling model is only faster than on the UCC model for small values of $p$,
after this point the value of $\rho_{ncc}$ is so small that its effects are no longer apparent.
While the results for the Email backbone indicate that correlations among neighbours are 
not the only influence on behaviour, it is clear that their presence does act to speed up 
diffusion processes over the temporal network.

\section{Conclusions}

The influence that memory in temporal networks has \cg{on the link
dynamics of the networks themselves as well as} on processes that run
on them is increasingly seen as key to our understanding of the way
that our highly networked world operates. In the context of spreading
processes on temporal networks, such as the passage of infections or
the diffusion of information, it has been observed that the presence
of memory can either speed up or slow down the spreading relative to
some memoryless case.  This result has been observed here in a manner
reminiscent of other recent findings. What is generally less well
understood is precisely how memory causes this change in the speed of
spreading \cg{and the role of the multivariate structure of interactions
of the link dynamics}. A great deal of work has gone into the studying
of how the correlated bursts in link activity that are the result of
non-exponential inter link times, and hence memory, slow down
spreading processes. This however does not yet give us a full picture.

Here we have introduced a novel, flexible and controllable
generative model for temporal networks which allows for 
arbitrary backbone topologies, and precise control over the
memory strength, memory length, average degree, and 
coupling strength.
\cg{The new model can be interpreted as the temporal
generalisation of the Erd\H{o}s-Renyi random graph with non-Markovian
memory, backbone structure, and both self- and cross- interactions of
links.  Not only this, but the model is simple enough to allow
applications to empirical data by using maximum likelihood estimation
methods.}

\cg{Hence, first of all, we have empirically proved that, despite the
  simplicity of the model, it is possible to infer many memory
  patterns observed in the real world. By estimating the model on a
  number of datasets, we have shown that: transportation networks are
  Markovian systems with significant cross-correlations of links
  capturing the presence of transport connections between the physical
  nodes, \ie stops, of the network; social networks (both online and
  offline) are in general non-Markovian with a memory order larger
  than one, and display auto-correlations of links, as a result of the
  stability pattern characterising some social ties, such as
  friendship; football networks are Markovian and display both
  link-specific persistence and cross interactions between links, with
  the two memory patterns which are inversely correlated as functions
  of the time resolution, as a result of the underlying contact
  dynamics of the game. Finally, within our framework, we have further
  validated the importance of describing the cross-interactions of
  links with a study of link prediction, thus showing how we are able
  to improve significantly the forecasting performance when such
  effects are included.}

The model is rich but flexible enough to allow the analytical treatment of a number of theoretical problems,
including finding the exact correlations between any two links.
Given this we have been able to study exactly how the memory
and coupling of the dynamics of different links 
in our model influence a spreading processes on the network. 
In doing this we have provided a solution to the time taken for 
a diffusion process to reach equilibrium in the limit of no cross
correlations and as a 
function of the memory length $p$. This shows us that the spreading time
is non-monotonically dependent on $p$, and allows us to infer that the 
equivalent memoryless process provides the fastest possible 
diffusion in our model. Looking at networks we have 
shown that correlations play a subtler role than might
previously been expected. While we find, in accordance with
previous works, that non-exponentially decaying autocorrelations 
among links do slow down diffusion, we, surprisingly, see that the opposite is
true of local correlations. When links that share a node are 
correlated, this tends to speed up diffusion. This is made clear 
by the fact that when we observe a system in which links have
fixed autocorrelation, but the correlations between neighbouring links 
varies (while all other parameters are kept constant), then diffusion is faster 
when correlations among neighbours are higher.
This has strong implications for real world systems. 
While it is understood that memory and correlations between links
have an effect on the spreading of information, the observation that 
correlations between neighbours and autocorrelations behave in 
opposite ways directly contributes to our understanding of many 
empirical systems. For example, when considering the diffusion of information
over a social network, and any consequent formation of opinions,
correlations between two different social ties must be considered as important
as the correlation of a social tie with its own history.
In a more general sense our findings also suggest that 
considering the evolution of links as independent processes in a temporal network
means we loose a significant amount of information.
Hence, when assessing the properties of an empirical network, correlations between 
the evolutions of links must be taken into account.
Finally, we have been able to test our model using as backbones the topologies 
of real-world systems. 
The differences in spreading behaviour demonstrated among these backbones show the important role that 
such topologies play. It is clear that memory, correlations and backbone
interact in a complex manner, and when considering the study of 
real world systems one can not assume to study of any of these 
features in isolation. Here, however, we have provided a 
framework in which the interplay between these features can be studied
systematically, and how surprising results occur when we do.

\begin{acknowledgments}
VL work was funded by the Leverhulme Trust Research Fellowship
“CREATE: the network components of creativity and success”. FL and PM acknowledge support by the European Integrated Infrastructure for Social Mining and Big Data Analytics (SoBigData++, Grant Agreement \#871042).
We also thank Lucas Lacasa for useful discussions.

\end{acknowledgments}

\section*{Author contributions}
All the authors designed the research. O.W. \cg{and P.M.}
performed calculations and generated figures.
All the authors discussed intermediate and final results and wrote the paper.

\section*{Competing interests}
The authors declare no competing interests, financial or otherwise.

\section*{Correspondance}
Correspondence and requests for materials should be addressed to O.W or V.L.

\section*{Open-source software}
The codes for both simulation and estimation of the CDARN(p) models proposed in this paper are available online at \href{http://www.sobigdata.eu/}{http://www.sobigdata.eu/}.

\section{Appendix}

\subsection{Linear indexing for links}

Throughout this work we frequently make use of a linear indexing for
links in a network, and, equivalently, entries in a matrix. In 
practice this is a way to take a pair of indices for either network nodes, or
rows and columns of the adjacency matrix, say $(i,j)$, and map 
it to a single number $\ell$. As only requirement we need for such a
mapping to be bijective, that is each unique pair $(i,j)$
corresponds to a unique value $\ell$. 
As an example, the simplest way this can be done, assuming 
$i,j \in \{1,...,N\}$ for some value $N$, is to take 
$\ell = N(i-1) + j$ 
element of a matrix as 1, then proceeding left to right
line by line.

\subsection{Average degree of the CDARN($p$) network}

Our aim is to show that a CDARN($p$) network on a complete backbone 
has the same average degree as a DARN($p$) network, and hence as an
Erd\H{o}s-Renyi (ER) random graph.  To do this we need only show that
the average value of an arbitrary link is given by $\left< a^{ij}_t
\right> = y$. Let us proceed by first averaging over the left and
right hand sides of Eq.~\ref{CDARN_rv_eqn} to get
\begin{equation}
	\left< a^{ij}_t \right> = q \left< a_{(t-Z^{ij}_t )}^{M^{ij}_t} \right> + (1-q)y,
\end{equation}
where the symbols $< \cdot >$ denote temporal averages. 
If we now label the link $(i,j)$ with its linear index $\ell =1,2,\ldots, L$, we
obtain the following
\begin{eqnarray}
\begin{aligned}
	\left< a^{\ell}_t \right> =& \frac{q}{p} \sum_{s=1}^{p} \sum_{\ell'=1}^{L}  c^{\ell \ell'} \left<a^{\ell'}_{t-s} \right> + (1-q) y, \\
	=& q \sum_{\ell'=1}^{L} c^{\ell \ell'} \left< a^{\ell'}_t \right> + (1-q)y.
\end{aligned}
\end{eqnarray}
where $c^{\ell \ell'}$ are the entries of the coupling matrix $ \underline{\underline{C}}$. 
In the above we have made use of the stationarity of the sequence $a^{\ell}_{t}$ 
to say that $\left< a^{\ell}_{t-a} \right> =
\left< a^{\ell}_{t} \right>$.  One can see also that $\left<
a^{\ell}_{t} \right> = \bar{a}$, for some constant $\bar{a}$, is a
solution to the above equations. The fact that $ \underline{\underline{C}}$ is row stochastic,
and so its rows sum to 1, then gives us that $\bar{a} = y$ is the
unique solution. Hence, we have obtained
$\left< a^{\ell}_{t} \right> = y$, that is we have shown that the 
CDARN($p$) produces networks with the same average degree of the
DARN($p$) model.

\subsection{The infinite memory limit} 
One of the key features of the DARN($p$) model 
 is that, as $p \to \infty$, it produces temporal networks that are indistinguishable 
from a sequence of independent ER graphs. 
Our aim now is to show that this is true for the CDARN($p$) model as well. 
We start by writing the 
conditional probability for a single link with linear index $\ell$:
\begin{equation}
	\text{Prob} \left( a^{\ell}_t = 1 | \{\underline{\underline{A}}_{s}\}_{s=t-1}^{t-p} \right) = (1-q)y + q \phi_t (p),
\end{equation}
where $ \{\underline{\underline{A}}_{s}\}$ is the random matrix representing the adjacency matrix 
at time $s$, which we say has observed values $a^{\ell}_{s}$.
We hence see that our problem can be reduced to a study of the
properties of some kernel function $\phi$, defined as
the probability that a 1 is drawn from any point in the memory, i.e.
\begin{equation}
	\phi_t (p) = \sum_{\ell'} c^{\ell \ell'} \frac{1}{p}  \sum_{k=1}^{p} a^{\ell'}_{t-k}.
\end{equation}
We recognise the sample expectation over the past $p$ steps of the time series, and so can see that
$\phi_t (p) \to y$ as $p \to \infty$. For completeness we must also check that
any fluctuations away from the mean can be ignored at finite times. 
First we see the following
\begin{eqnarray}
\begin{aligned}
	\phi_{t+1}(p) - \phi_t(p) =& \sum_{\ell'} c^{\ell \ell'} \frac{1}{p} \sum_{k=1}^p \left( a^{\ell'}_{t-k+1} - a^{\ell'}_{t-k} \right), \\
	=& \frac{1}{p} \sum_{\ell'} c^{\ell \ell'}  \left( a^{\ell'}_{t-k+1} - a^{\ell'}_{t-k} \right).
\end{aligned}
\end{eqnarray}
We then have that $a^{\ell'}_{t-k+1} - a^{\ell'}_{t-k} \in \{ -1,0,1 \}$, and so
\begin{eqnarray}
	-\frac{1}{p} \leq \phi_{t+1}(p) - \phi_t(p) \leq \frac{1}{p}, \\
	\implies y - \frac{t}{p} \leq \phi_t(p) \leq y+ \frac{t}{p}.
\end{eqnarray}
Hence, in the large $p$ limit then the memory kernel $\phi$ tends to 0, and
so the system is equivalent to one in which there is no memory. 
This displays exactly the same behaviour as is found for the DARN($p$) model: 
the CDARN($p$) model does indeed tend to a memoryless 
model in the limit of large memory.
In the memoryless case we note that $\text{Prob} \left( a^{\ell}_t = 1\right) = y$, and so
must have an expected inter-link time of $1/y$, and correspondingly the 
expected time until the  $n-$th  link is $n/y$.

\subsection{Time to equilibrium in a two node system with a permanent link}
Consider the equations defining diffusion in continuous time between two nodes,
for which the link between them is permanent (always present):
\begin{equation}
	\dot{d}^1(t) = - \mu \left( d^1(t) - d^2(t) \right).
\end{equation}
If we impose conservation, i.e. $d^2(t) = 1 - d^1(t)$, (and 
drop the 1 so that $d^1(t) \to d(t)$) we can rewrite this as
\begin{equation}
	\dot{d}(t) = \mu - 2 \mu d(t).
\end{equation}
Assuming $d(0)=1$, the solution is
\begin{equation}
	d(t) = \frac{1}{2} \left( e^{-2\mu t} + 1 \right).
	\label{eqn:two_node_soln}
\end{equation}
We say that this system has reached equilibrium at the first time
(in a continuous sense) $t = \tau_c$ where
$\left| d^1(t) - d^2(t) \right| < \epsilon $ for some small positive 
$\epsilon$. Again, imposing conservation this can be rewritten as
the first value of $t$ such that $2d(t) - 1 = \epsilon$.
With Eq.~\ref{eqn:two_node_soln} we can then find the (continuous) time to equilibrium
directly as
\begin{equation}
	\tau_c = \frac{- \log \epsilon}{2 \mu}.
\end{equation}
Hence the number of full time steps $\tau$ of length $\Delta t$ which must occur
before equilibrium is reached is given by
\begin{equation}
	\tau = \floor{ \frac{- \log \epsilon}{2 \mu \Delta t}}.
	\label{steps_to_eqm}
\end{equation}
Note that for this system, for any given values of $\mu$ and $\Delta t$ 
we can always find $\bar{\mu} = \mu \Delta t$, meaning that we may 
fix $\Delta t = 1$ and still recover the full range of possible values for 
$\bar{\mu}$ by varying $\mu$.

\subsection{The transition matrix for a DAR($p$) variable}
Consider a stochastic process where the random variable $X_t$ is
governed by the the DAR($p$) model:
\begin{equation}
  X_t = Q_t X_{(t-Z_t )} + (1-Q_t)Y_t 
 \label{SI:DARp_eqn}
\end{equation}
where, for each $t$, $Q_t \sim \mathcal{B}(q)$ and $Y_t \sim \mathcal{B}(y)$ are Bernoulli random variables, while $Z_t$ 
 picks integers uniformly from the set $\{1,...,p\}$. This can be thought of as
 a $p$-th order Markov chain, and so is equivalent to a first order Markov chain 
 in an enlarged state space \cite{macdonald1997hidden}. Accordingly,
 we define the so-called ``$p$-state'' of link $(i,j)$ at time $t$, by combining the state of
the link at time $t$ along with its previous $p-1$ states in the
vector $\underline{S}_t= \left(X_t,X_{t-1},...,X_{t-p+1}\right)$.
If we now define the set $\mathcal{S}$ as the set containing 
all $2^p$ possible $p$-states, then for any $\alpha,\beta \in \mathcal{S}$ 
we can look at the conditional probability 
${\rm Prob}(\underline{S}_{t+1} = \beta | \underline{S}_t= \alpha)$.
This defines the entries $T_{\alpha \beta}$ of the $p$-th order
$2^p\times2^p$ transition matrix. More details on the transition
matrix of the DAR($p$) model can be found in \cite{Williams_2019}.

\subsection{State indexing}

When writing the matrix element $T_{\alpha \beta}$ we are implicitly
associating an index to the $p-$states $\alpha$ and $\beta$.
Since elements of a matrix are usually labeled by values $i,j \in \{0,...,I-1\}$
(or $i,j \in \{1,...,I\}$) for some value of $I$, 
we must hence impose an ordering on the states $\alpha,\beta \in \mathcal{S}$.
This is done by associating a linear index $l(\alpha)$ to each possible state $\alpha \in \mathcal{S}$ (and similarly for $\beta$).
The simplest form
of this labelling function in our case, given a memory length of $p$, is 
\begin{equation}	
	l(\alpha) = \sum_{k=0}^{p} 2^k  \alpha_k,
	\label{eqn:labeling}
\end{equation}
where $\alpha_k$ is the $k_{th}$ entry in the $p$-state vector associated with $\alpha$.
In practice this is equivalent to consider the sequence of 0's and 1's,  
representing the link history corresponding to state $\alpha$, as a binary
number and converting in into to a decimal number.
We will implicitly assume that wherever we use $\alpha$,
or any state in $\mathcal{S}$, we are referring to the label $l(\alpha)$,
and that the labelling function is as given in Eq.~\ref{eqn:labeling}.

\subsection{\cg{Initialising CDARN($p$) model simulations}}

\cg{When generating realisations of the CDARN($p$) model for the purposes of 
Monte-Carlo simulation (or any other simulation) it is important to ensure that the 
model is appropriately initialised. In all of the simulations and calculations here 
we require that the model be in a steady state, and so before any simulated diffusion starts
we do the following: 
\begin{itemize}
\item For each link $\ell \in \{1,...,L\}$ and for each time $s \in \{0,..., p-1\}$ assign to link state $a_{s}^{\ell}$ the value taken by a random variable $X_{s}^{\ell} \sim \mathcal{B}(y)$. This gives us a set of pre-initial conditions from which simulation can be started. \\
\item Simulate the CDARN($p$) model for times $p,...., T_0$ using the previously generated states of the network, for some large $T_0$. $T_0$ is chosen so that the network has reached a steady state, as approximated by the point where the autocorrelations $\left<  a_{p}^{\ell} a_{T_0}^{\ell} \right>$ have decayed below some suitable threshold. Functionally this has been set at approximately $T_0 = 500$.
\item Any simulation on top of the network can now start. This process must be repeated for every simulated realisation. 
\end{itemize}
} 

\subsection{Average time to equilibrium for a single Markov link}

The value $\left< \tau^1 \right>$ of the average time until  
diffusion across a single DAR(1) link reaches equilibrium,
is central to any analysis of the rescaled times  
to equilibrium.  Hence, we calculate it explicitly here. We know that,
given the value $\tau = n$ from Eq.~\ref{steps_to_eqm} giving the
number of time steps until equilibrium in the two node system where the link is always present, 
$\left< \tau^1 \right>$ will be precisely the time taken for a DAR(1) link to occur $n$ times.
This can be found as the solution to the following equation: 
\begin{equation}
	\left< \tau^1 \right> = \underline{\omega}^T \left( \sum_{t=0}^{n-1} \underline{\underline{h}}^t \right) \underline{k},
\end{equation}
where $\underline{\underline{h}}^t$ denotes the $t_{th}$ power of the matrix $\underline{\underline{h}}$.
Now $\underline{\underline{h}}$ will be a $2 \times 2$ matrix, and
$\underline{\omega}$ and $\underline{k}$ will be two dimensional
vectors.
%
Given the definition of $h_{\alpha \beta}$ as the probability that a system starting in state $\alpha$ 
ends in state $\beta$, we can easily see that the following must be true:
\begin{equation}
	\underline{\underline{h}} = 
	\begin{pmatrix}
		0 & 1 \\
		0 & 1
	\end{pmatrix},
\end{equation}
and so
\begin{equation}
	\sum_{t=0}^{n-1} \underline{\underline{h}}^t = 
	\begin{pmatrix}
		1 & n-1 \\
		0 & n
	\end{pmatrix}.
\end{equation}

Similarly we may find that $\omega_{1} = 1-y$ and $\omega_{2} = y$, 
and $k_1 = \left( \left(1-q \right) y \right)^{-1}$ and $k_{2} = y^{-1}$.
Giving us the equation
\begin{equation}
	\left< \tau^1 \right> = \left( \frac{1}{1-q} + n - 1 \right) \frac{1-y}{y} + n.
\end{equation}

It is then a simple matter to extend this result and calculate the
value of the rescaled time to equilibrium $\mathfrak{T}^{1}$
 directly. Given that we know $\left< \tau^p \right> \to n/y$ 
 as $p \to \infty$, and this is precisely the value of $\left< \tau^0 \right>$,
 we then find: 
 \begin{equation}
   \mathfrak{T}^{1} = \frac{\left< \tau^1 \right>}{\left<\tau^{0} \right>} = \left( \frac{1}{1-q} + n - 1 \right) \frac{1-y}{n} + y.
\end{equation}
It is easy to see that the maxima and minima of this function 
in terms of $n$ and $y$ are finite and occur at their limiting 
values ($y = 0,1$ and $n = 1,\infty$ respectively) if $q \neq 1$.
However in the limit $q \to 1$ we see that $\mathfrak{T}^{1} \to \infty$. In the 
$q=0$ limit we obtain the value $\mathfrak{T}^{\infty} = 1$, as expected.

\subsection{Rescaled time to equilibrium in the limit of large $p$}

In the main text we claim that $\mathfrak{T}^{p} \geq 1$ for suitably
sparse initial conditions, i.e. when $y$ is small,  
but that for $y \approx 1$
the opposite can be true. 
To understand this we first formalise our statement: 
given an initial probability vector $\underline{\omega}$, we have 
 $ \left<\tau^{p} \right>  \geq  \left<\tau^{\infty} \right> $, 
 provided that the entries representing states in which no link is present
 (in our case $\omega_{\alpha} $ for $\alpha \in \{0,...,2^{p-1} \}$) contain the majority 
 of the probability mass.
 Recall first that we are implicitly labelling our states $\alpha \in \mathcal{S}$
 according to the labelling function given in Eq.~\ref{eqn:labeling}.
Now, notice that, since $\underline{\omega}$ is a probability vector, 
and $\underline{\underline{h}}$ is a stochastic matrix, we can define 
a vector $\underline{k}^{0}$ such that $k^{0}_{\alpha} = 1/y$ for all $\alpha$, and we 
can write the following equation:
\begin{equation}
	\left<\tau^{\infty} \right> =  \underline{\omega}^T \left( \sum_{t=0}^{n-1} \underline{\underline{h}}^t \right) \underline{k}^{0}.
\end{equation}
Hence we can write
\begin{equation}
	\left<\tau^{p} \right> - \left<\tau^{\infty} \right> =  \underline{\omega}^T \left( \sum_{t=0}^{n-1} \underline{\underline{h}}^t  \left( \underline{k} - \underline{k}^{0} \right) \right).
\end{equation}
By construction $h_{\alpha \beta} = 0 $ if $\beta < 2^{p-1}$, and so if we define 
\begin{equation}
	\underline{\tilde{\omega}}^T = \underline{\omega}^T   \left( \sum_{t=0}^{n-1} \underline{\underline{h}}^t \right),
\end{equation}
then, when $\beta < 2^{p-1}$, we have $\tilde{\omega}_{\beta} = \omega_{\beta}$.
Hence, we see that if $\omega_0 \approx 1$ (the entry in $\omega$ 
representing an initial state with no links), then we need only check
that $k_0 > 1/y$ to show that $\mathfrak{T}^{p} \geq 1$.
This can be checked directly by analysing
the average time taken to reach equilibrium $k_{\alpha}$, given some starting state 
$\alpha$, as defined by the following set of linear equations:
\begin{equation}
	k_{\alpha} = 1 + T_{\alpha \alpha'}k_{\alpha'},
\label{avg_hit_lin_form}
\end{equation}
where $\alpha' = \floor{\alpha/2}$.
From this we can directly obtain
\begin{equation}
	k_0 = \frac{1-q p^{-1}}{(1-q)y},
\end{equation}
and hence confirm that $k_0 > 1/y$ when $\omega_0 \approx 1$.
We now want to understand the conditions in which this breaks down, and we instead observe $\mathfrak{T}^{p} < 1$. 
One can manually check that, for any values of $p$ or $q$, $k_{\alpha} > 1/y$ for $\alpha=0,1$, but that
this inequality does not generally hold for $\alpha=4$. 
As a specific example of this, if we fix $p=3$, $y=0.01$ and $q=0.1$, then $1/y = 100$, but 
$k_4 \approx 98.52$.
To understand this behaviour, we can then make
use of the following two facts about $k_{\alpha}$. 
Given a memory state $\alpha \in \mathcal{S}$,
\begin{itemize}
\item If by $\alpha_n$ we indicate
  the memory state with a 1 in the $n-$th entry,
  and zeros elsewhere, then the values of $k_{\alpha_n}$ are given by
  solutions to the equation $x_{n+1} = 1 + ax_{n}$, with appropriate
  values for $a$ and $x_1$.
\item if $\beta$ is the memory state obtained by taking memory state
  $\alpha$ and replacing any of its 0 states with 1, then $k_{\beta} <
  k_{\alpha}$.
\end{itemize}  
To prove the first of these statements, we directly analyse Eq.~\ref{avg_hit_lin_form}.
This equation, in our $\alpha$ notation becomes $k_{\alpha_{n+1}} = 1 + T_{\alpha_{n+1} , \alpha_{n}} k_{\alpha_n}$,
but we also notice that $T_{\alpha_{n+1} , \alpha_{n}}$ is invariant of $n$, always taking the value
$T_{\alpha_{n+1} , \alpha_{n}} = 1 - q/p - (1-q)y$, which we will now denote as $T$. To obtain the desired form
of difference equation, we now simply identify $x_n = k_{\alpha_n}$ and $a = T$.
This can be easily solved to give the following:
\begin{equation}
	k_{\alpha_n} = \frac{T^n}{(1-q)y} + \frac{1-T^n}{1-T}.
\end{equation}
This equation must clearly be decreasing with n.\\

To prove the second of these statements
consider two possible memory states $\alpha$ and $\beta$, 
where $\beta$ is given by taking $\alpha$ and replacing one of the
zeros in its memory with a one. Let us label the position of the 
state which we change with $t$. Now let us now denote
$\alpha^{(n)} = \floor{\alpha^{(n-1)}/2}$, 
where $\alpha^{0} = \alpha$, and similarly for $\beta$. 
To clarify, we can think of $\alpha^{(n)}$ as being the memory state $\alpha$ shifted 
back $n$ times, or similarly what happens to the memory state of a DAR($p$) process
if it starts in state $\alpha$ and generates $n$ zeros.
We can then write the
following equation directly from Eq.~\ref{avg_hit_lin_form}:
\begin{equation}
	k_{\alpha^{(n)}} - k_{\beta^{(n)}} = T_{\alpha^{(n)} \alpha^{(n+1)}} k_{\alpha^{(n+1)}} - T_{\beta^{(n)} \beta^{(n+1)}} k_{\beta^{(n+1)}}.
\end{equation}
Given our definition in Eq.~\ref{transition_mat},
and the fact that $\beta$ is $\alpha$ with a 1 added, we can rearrange this to give
\begin{equation}
	k_{\alpha^{(n)}} - k_{\beta^{(n)}}  = T_{\alpha^{(n)} \alpha^{(n+1)}} \left(k_{\alpha^{(n+1)}} - k_{\beta^{(n+1)}} \right) + \frac{q}{p} k_{\beta^{(n+1)}}.
\end{equation}
From this we can see that if $k_{\alpha^{(n+1)}} \geq k_{\beta^{(n+1)}}$ 
then we must have $k_{\alpha^{(n)}} \geq k_{\beta^{(n)}}$. Now, by construction 
we know that $\alpha^{(n)} = \beta^{(n)} ~ \forall n \geq t$, since
this is the point at which the additional 1 in the memory is removed.
In turn this means that $k_{\alpha^{(n)}} = k_{\beta^{(n)}} ~\forall n \geq t$. 
Inductively this gives us 
that 
\begin{equation}
	k_{\alpha^{(t-1)}} \geq k_{\beta^{(t-1)}},...,k_{\alpha^{(n+1)}} \geq k_{\beta^{(n+1)}}.
\end{equation}
Thus we have that $k_{\alpha^{(n)}} \geq k_{\beta^{(n)}}$. 
Hence we have proved that 
$k_{\alpha}$ is decreased by adding a one at any point in the memory,
and, equally, increased by adding a zero at any point in the memory.
\\
The first statement gives us that, since we can not guarantee that 
$k_4 > 1/y$, we can not guarantee that, for any state $\alpha$ with 
a single one in any position other than 1, $k_{\alpha} > 1/y$.
The second statement then tells us that, since any state 
$\alpha$ can be generated by taking a state with only a single 1 somewhere,
and adding more 1's to it, 
we can never guarantee that $k_{\alpha} > 1/y$ for any $\alpha \geq 4$.  

Because of this we see that for small $y$ we must have 
$\left< \tau^p \right> \geq \left< \tau^{\infty} \right>$, and hence
we must have $\left< \tau^p \right> \geq \left< \tau^{0} \right>$, finally
giving us that $\mathfrak{T}^p \geq 1$. However, for larger $y$ this may not be the case.

\subsection{Correlations in the CDARN($p$) model}

By introducing the possibility that a link in a DARN($p$) network can
draw from the memory of another link, and hence creating the
CDARN($p$) model, we have introduced correlations among the activities
of different links. As we will show in this appendix,  
the extent of these correlations can be completely 
characterised analytically. If we have a network with $L$ possible 
links, each with its own linear index, let us denote the correlations 
between link $\ell$ and $\ell'$ at time lag $k$ as $\left< a^{\ell}_t a^{\ell'}_{t-k} \right> = \rho^{\ell \ell'}_{k}$. 
Following the procedures in \cite{Williams_2019,jacobs1978discrete}, we can 
derive the Yule-Walker equations: 
\begin{equation}
	\rho^{\ell \ell'}_k = \frac{q}{p} \sum_{a=1}^{p} \sum_{b=1}^{L} c^{\ell b} \rho^{b\ell'}_{k-a},
\end{equation}
where the elements $c^{\ell b}$ are taken from the coupling matrix assigning 
the probability of a link $\ell$ drawing from the memory of link $b$.
This can be written more compactly in terms of the corresponding 
matrices $ \underline{\underline{\rho}}_k = \{ \rho^{\ell \ell'}_k  \}$ and $ \underline{\underline{C}} = \{ c^{\ell \ell'} \}$ as 
\begin{equation}
	\underline{\underline{\rho}}_k = \frac{q}{p} \underline{\underline{C}} \sum_{a=1}^{p}  \underline{\underline{\rho}}_{k-a}.
	\label{corr_yule_walker}
\end{equation} 
These equations can be solved given a suitable closure.
Following \cite{Williams_2019}, we can re-write this 
expression for values of $k < p$ as 
\begin{equation}
	\underline{\underline{\rho}}_k = \frac{q}{p}  \underline{\underline{C}} \left( \sum_{a=1}^{k-1}  \underline{\underline{\rho}}_a +  \sum_{a=1}^{p-k}  \underline{\underline{\rho}}_a +  \underline{\underline{\rho}}_0 \right),
	\label{gen_close}
\end{equation}
for some value $ \underline{\underline{\rho}}_0$.
This equation can be seen to have a constant solution $ \underline{\underline{\rho}}$, which satisfies: 
\begin{equation}
	 \underline{\underline{\rho}} = \frac{q}{p}  \underline{\underline{C}} \left( (p-1)  \underline{\underline{\rho}} +  \underline{\underline{\rho}}_0 \right).
	\label{eqn:gen_rho}
\end{equation}
Now we need to find a suitable expression for $ \underline{\underline{\rho}}_0$.
We know that, by definition, $\rho^{\ell \ell}_0 = 1$. The off diagonal 
entries however are given by the Yule-Walker equation 
\begin{equation}
	\rho^{\ell \ell'}_0 = \frac{q}{p} \sum_{a=1}^{p} \sum_{b=1}^{L} c^{\ell b} \rho^{b \ell'}_{a}.
\end{equation}
But, we know that the value of $ \underline{\underline{\rho}}_a$ must be a constant $ \underline{\underline{\rho}}$, and so the 
off-diagonal elements of $ \underline{\underline{\rho}}_0$ will be the same as the off-diagonal 
elements of 
\begin{eqnarray}
\begin{aligned}
	 \underline{\underline{\bar{\rho}}}_0 =& \frac{q}{p} \sum_{a=1}^p  \underline{\underline{C}}  \underline{\underline{\rho}},\\
	=& q  \underline{\underline{C}}  \underline{\underline{\rho}}.
	\label{gen_rho_zero}
\end{aligned}
\end{eqnarray}
Putting everything together we get the equation
\begin{equation}
	\rho^{\ell \ell'} = \frac{q}{p} \left( (p-1) \sum_{b=1}^{L} c^{\ell b} \rho^{b \ell'} + q \sum_{b \neq \ell'} \sum_{\ell''=1}^{L} c^{\ell b} c^{b \ell''} \rho^{\ell'' \ell'} + c^{\ell \ell'} \right).
\end{equation}
This can be rearranged to give
\begin{equation}
	\rho^{\ell \ell'} = \frac{q}{p} \left( \sum_{\ell''=1}^{L} \left( (p-1) c^{\ell \ell''} + q \sum_{b \neq \ell'} c^{\ell b} c^{b\ell''} \right) \rho^{\ell'' \ell'} + c^{\ell \ell'} \right).
\end{equation}
This can be further simplified by constructing the tensor $ \underline{\underline{\underline{\Delta}}}$ as
 \begin{equation}
	\Delta^{\ell \ell' \ell''} = \frac{q}{p} \left( (p-1) c^{\ell \ell''} + q \sum_{b \neq \ell'} c^{\ell b} c^{b \ell''} \right),
\end{equation}
The system of equations given in Eq.~\ref{eqn:gen_rho} can then
be written as
\begin{equation}
	\rho^{\ell \ell'} = \sum_{\ell''=1}^{L} \Delta^{\ell \ell' \ell''} \rho^{\ell'' \ell'} +  \frac{q}{p} c^{\ell \ell''}.
	\label{gen_rho_full_eqn}
\end{equation}
 This form is more easily dealt with in numerical applications, as 
 a simple dimensional reduction (flattening) yields a more 
 traditional form for a system of linear equations.
 Importantly, this solution relies on no 
 properties of the coupling matrix other than stochasticity, 
 which it must have by definition. In special cases, such as those
 where coupling is uniform or symmetric, we can simplify these equations
 further by analysing the symmetries that arise in $ \underline{\underline{C}}$ and $ \underline{\underline{\underline{\Delta}}}$.

\subsection{Evolution of the autocorrelation function}
We wish to now show that the full extent of the autocorrelations
in our model are described by the constant value $ \underline{\underline{\rho}}$, given over the 
first $p$ time steps. 
We first notice that from Eq.~\ref{corr_yule_walker} we can obtain the following:
\begin{eqnarray}
\begin{aligned}
	 \underline{\underline{\rho}}_k -  \underline{\underline{\rho}}_{k-1} =& \frac{q}{p}  \underline{\underline{C}} \left( \sum_{t=1}^{p}  \underline{\underline{\rho}}_{k-t} - \sum_{t=1}^{p}  \underline{\underline{\rho}}_{k-t-1} \right), \\
	=& \frac{q}{p}  \underline{\underline{C}} \left(  \underline{\underline{\rho}}_{k-1} -  \underline{\underline{\rho}}_{k-p-1} \right),
\end{aligned}
\end{eqnarray}
and hence
\begin{equation}
	 \underline{\underline{\rho}}_k - \left(  \underline{\underline{I_d}}+ \frac{q}{p} \underline{\underline{C}} \right)  \underline{\underline{\rho}}_{k-1} = -\frac{q}{p}  \underline{\underline{C}}  \underline{\underline{\rho}}_{k-p-1}.
	\label{rho_diff}
\end{equation} 
However, we know that for $k \in \{1,...,p\}$ the autocorrelation is a constant $ \underline{\underline{\rho}}_k =  \underline{\underline{\rho}}$,
meaning that for $k \in \{p+1,...,2p+1\}$ Eq.~\ref{rho_diff} becomes
\begin{equation}
	 \underline{\underline{\rho}}_k - \left(  \underline{\underline{I_d}}+ \frac{q}{p}  \underline{\underline{C}} \right)  \underline{\underline{\rho}}_{k-1} = - \frac{q}{p}  \underline{\underline{C}}  \underline{\underline{\rho}}.
\end{equation}
 This is now a first order inhomogeneous difference equation with solution
 \begin{equation}
	 \underline{\underline{\rho}}_k =  \underline{\underline{\rho}} -  e^{k \log \left(1 + \frac{q}{p}  \underline{\underline{C}} \right)}  \underline{\underline{R}},
\end{equation}
where $ \underline{\underline{R}}$ is a constant matrix. By noticing that 
$ \underline{\underline{\rho}}_{p+1} = q  \underline{\underline{C}}  \underline{\underline{\rho}}$ we obtain the expression
\begin{equation}
	 \underline{\underline{R}} = \left(1-q \right)  \underline{\underline{\rho}} \left( 1 + \frac{q}{p} \right)^{-\left(p+1\right)}.
\end{equation} 
 With this solution we see that the equation governing the values of $ \underline{\underline{\rho}}_k$, for $k$ in the range
 $p+1$ to $2p+1$, 
 is of the form
 \begin{equation}
 	 \underline{\underline{\rho}}_k - \bar{q}  \underline{\underline{\rho}}_{k-1} = - \underline{\underline{A}} + e^{- \underline{\underline{\lambda}} k} \underline{\underline{B}},
\end{equation}
where $\bar{q}$, $ \underline{\underline{A}}$, $ \underline{\underline{B}}$ and $ \underline{\underline{\lambda}}$ are constant matrices. 
This equations has a general solution 
\begin{equation}
	 \underline{\underline{\rho}}_k =  \underline{\underline{A}}' - e^{- \underline{\underline{\lambda}} k}  \underline{\underline{B}}'.
\end{equation}
where $ \underline{\underline{A}}'$ is a constant matrix, and 
$\underline{\underline{B}}'$ is a matrix that is constant over every 
interval $k \in [np + 1, (n+1)p + 1]$. 
Moreover, in our specific case we find that $ \underline{\underline{A}}' =  \underline{\underline{\rho}}$ and $ \underline{\underline{\lambda}} = \log \left(1 + \frac{q}{p}  \underline{\underline{C}} \right)$.
This implies that not only is the autocorrelation function for the CDARN($p$) process 
exponentially decreasing for all values of $k$ larger than $p+1$, but also that
this decay varies according to a single parameter $ \underline{\underline{B}}'$ every $p$ time steps.
This gives us a full picture of the autocorrelations for a CDARN($p$) process.

\subsection{Special case: totally symmetric cross correlation}

The simplest type of correlation in the CDARN($p$) model occurs when
the coupling matrix $ \underline{\underline{C}}$ is such that $c^{\ell \ell'} = 1/L ~ \forall
\ell, \ell'$, i.e. regardless of the pair of links in question.
We can immediately notice that our tensor $ \underline{\underline{\underline{\Delta}}}$ now takes the form: 
\begin{equation}
	\Delta^{\ell \ell' \ell''} = \frac{q}{p} \left( (p-1) \frac{1}{L} + q \frac{L-1}{L^2} \right),
\end{equation}
which is invariant over the three indexes $\ell,\ell'$ and $\ell''$.
Consequently $\rho^{\ell \ell'}$ 
must be invariant over $\ell$ and $\ell'$. Hence all of the lagged correlations have
the same value, and we can write $\rho^{\ell \ell'} = \rho$ and $\Delta^{\ell \ell' \ell''} = \Delta$, giving us the following equation:
\begin{equation}
	\rho = L \Delta \rho + \frac{q}{Lp}.
\end{equation}
Solving for $\rho$ gives
\begin{equation}
	\rho = \left( Lp \left( \frac{1}{q} - 1 \right) + (1-q)L + q \right)^{-1}.
\label{eqn:symm_rho}
\end{equation}
This provides a full picture of the lagged correlations present in the
system.  All that remains is to find the time 0 correlations
$\rho_{0}^{\ell \ell'}$ when $\ell \neq \ell'$. This can be done as
follows:
\begin{eqnarray}
\begin{aligned}
  \rho^{\ell \ell'}_{0} =&  \frac{q}{Lp} \sum_{a=1}^{p} \sum_{b=1}^{L} \rho^{b \ell'}_{a}
  \\
  =&   \frac{q}{Lp} L \sum_{a=1}^p \rho_{a}
  \\
	=&q \rho.
\end{aligned}
\end{eqnarray}
Where the last line is given by the fact that for $a = 1,...,p$ we have $\rho_a = \rho$, the 
constant value given in Eq.~\ref{eqn:symm_rho}. 
Hence when $\ell \neq \ell'$, $\rho^{\ell \ell'}_0 = q\rho$.

Note first that if $L=1$ then we recover the autocorrelation function of a DAR($p$) process.
Also note that as $L$ increases this value must decrease, meaning that for large 
networks both correlations and autocorrelations are removed, and so memory 
no longer has any effect on the evolution of the system.

\subsection{Special case: Uniform Cross Correlation (UCC)}

The second special case we will consider is that of uniform cross
correlation (UCC), as induced by a symmetric coupling matrix.
Specifically this means that we require that $ \underline{\underline{C}}$ be symmetric, with
$c^{\ell \ell} = 1-c$ for all values of $\ell$ and some given value of
$c$, and $c^{\ell \ell} = \bar{c}$ for $\ell \neq \ell'$ with $\bar{c}
= c/(L-1)$.  Going back to the general case in
Eq.~\ref{gen_rho_full_eqn} we notice that these conditions ensure that
$\rho^{\ell \ell}$ is invariant with respect to $\ell$,
and when $\ell \neq \ell'$
$\rho^{\ell \ell'}$ is invariant with respect to $\ell$ and $\ell'$.  This means
that all of the values of $\rho^{\ell \ell'}$ can be found as the
solutions to the two following equations (note that $\ell \neq \ell'$
is assumed here)
\begin{eqnarray}
\begin{aligned}
	\rho^{\ell \ell} =& \sum_{\ell'' \neq \ell} \Delta^{\ell \ell \ell''} \rho^{\ell'' \ell} + \Delta^{\ell \ell \ell} \rho^{\ell \ell} + \frac{q}{p} (1-c), \\
	\rho^{\ell \ell'} =& \sum_{\ell'' \neq \ell , \ell'} \Delta^{\ell \ell' \ell''} \rho^{\ell'' \ell'} + \Delta^{\ell \ell' \ell} \rho^{\ell \ell'} + \Delta^{\ell \ell' \ell'} \rho^{\ell' \ell'} + \frac{q}{p} \bar{c}.
\end{aligned}
\end{eqnarray}
Hence we need only find the relevant values of $\Delta$ to proceed.
Given the definition of $\Delta$ and $c$ and $\bar{c}$ we can find the following:
\begin{eqnarray}
\begin{aligned}
	\Delta^{\ell \ell \ell} =& \frac{q}{p} \left( (p-1) (1-c) + q (L-1) \bar{c}^2 \right) \eqqcolon \Delta^1, \\
	\Delta^{\ell \ell \ell''} =& \frac{q}{p} \left( (p-1) \bar{c} + q \bar{c} \left((1-c) + (L-2) \bar{c} \right) \right) \eqqcolon \Delta^2, \\
	\Delta^{\ell \ell' \ell} =& \frac{q}{p} \left( (p-1) (1-c) + q \left( (1-c)^2 + (L-2) \bar{c}^2 \right) \right) \\ 
	\eqqcolon& \Delta^3, \\
	\Delta^{\ell \ell' \ell'} =& \Delta^{\ell \ell \ell''} \eqqcolon \Delta^2, \\
	\Delta^{\ell \ell' \ell''} =& \frac{q}{p} \left( (p-1) \bar{c} + q \bar{c} \left(2(1-c) + (L-3) \bar{c} \right) \right) \eqqcolon \Delta^4.
\end{aligned}
\end{eqnarray}
Noticing that $\Delta^{\ell \ell \ell''}$ and $\Delta^{\ell \ell' \ell''}$ are invariant of $\ell, \ell'$ and $\ell''$ (down to 
excluded values) we then obtain the following pair of equations:
\begin{eqnarray}
\begin{aligned}
	\rho^{\ell \ell} =& (L-1) \Delta^2 \rho^{\ell \ell'} + \Delta^1 \rho^{\ell \ell} + \frac{q}{p} (1-c), \\
	\rho^{\ell \ell'} =& \left(\Delta^3 + (L-2) \Delta^4 \right) \rho^{\ell \ell'} + \Delta^2 \rho^{\ell \ell} + \frac{q}{p} \bar{c}.
\end{aligned}
\end{eqnarray}
This gives us a simple, solvable pair of equations. Note that 
while the full solution in terms of $q,p,c$ and $L$ is easy to obtain now, we will not write it down 
here due to its length.
To find the time 0 correlations we can use Eq.~\ref{gen_rho_zero} to obtain
\begin{equation}
	\rho_{0}^{\ell \ell'} = q \left( \left((1-c) + (L-2) \bar{c} \right) \rho^{\ell \ell'} + \bar{c} \rho^{\ell \ell} \right).
\end{equation}
Competing the description of the correlations for the UCC model.

\subsection{Uniform cross correlation in the large network limit}

The uniform cross correlation (UCC) model for CDARN($p$) networks
was introduced to evenly distribute any temporal cross correlations between links over 
the entire network. In doing this we minimise the cross correlations,  i.e. 
$\rho^{\ell \ell'}$ where $\ell \neq \ell'$, for fixed values of $p,q,y$ and $c$.
In turn this minimises the influence that 
any such cross correlations have on the diffusion process over the network.
What we now show is that when the 
backbone of the temporal network has a large number of links then the
UCC model is indistinguishable from a DARN($p$) model that has been 
restricted to the same backbone, and hence temporal cross correlations 
between links are completely removed. 

Consider a temporal network given by time varying adjacency matrix $ \underline{\underline{A}}_t$ with
observed value $\{a^{\ell}_{t}\}$, and
with $L$ links, generated by the 
CDARN($p$) model with link density $y$, memory strength $q$, memory length $p$ 
and coupling matrix $\underline{\underline{C}}$ as in the UCC case. 
The conditional probability of a link $\ell$ 
  occurring at time $t$, given
the past $p$ states of the network can be thought of in terms of
contributions from the memory of the link itself, the memory of all
other links, and some background contribution. This can hence be
written as follows 
\begin{equation}
	\text{Prob}(a^{\ell}_t | \{\underline{\underline{A}}_{s}\}_{s = t}^{t-p} ) = (1-q)y + q \left( (1-c) \phi_{\rm self} + c \phi_{\rm other} \right),
\end{equation}
where $\phi_{\rm self}$ and $\phi_{\rm other}$ represent the contributions to
the conditional probability $\text{Prob}(a^{\ell}_t | \{ \{\underline{\underline{A}}_{s}\}_{s = t}^{t-p} )$
from the past $p$ states of the link $\ell$ and every
other link respectively.\\ For links to be effectively independent
then we require that as $L \to \infty$, $\phi_{\rm other}$ tends to a
constant, and hence the link $\ell$ has no memory of the past states of
any other link. To show this we study the memory kernels $\phi_{\rm self}$
and $\phi_{\rm other}$ directly as:
\begin{eqnarray}
\begin{aligned}
	\phi_{\rm self} =&  \frac{(1-c)}{p} \sum_{k=1}^{p} a^{\ell}_{t-k},\\
	\phi_{\rm other} =& \frac{c}{(L-1)p} \sum_{\ell' \neq \ell} \sum_{k=1}^{p} a^{\ell'}_{t-k}.
\end{aligned}
\end{eqnarray}
We need only focus on $\phi_{\rm other}$.
First, let us consider the average value
$\left< a^{\ell'}_{t-k} \right>_{\ell'}$.
The CDARN($p$) network is taken to be in a stationary state, and so
the symmetry of the links under any relabelling guarantees us that
$\text{Prob}(a^{\ell'}_{t-k})$ is the same for
each link $\ell'$ and for each time $t-k$. Hence we can write $
\text{Prob}(a^{\ell'}_{t-k}) = \bar{a} ~ \forall \ell'$ for some constant $\bar{a}$. Then
we must have, for any of the $L-1$ possible values of $\ell'$,
\begin{equation}
	\left< a^{\ell'}_{t-k} \right>_{\ell'} = \text{Prob}(a^{\ell'}_{t-k}) = \bar{a}.
\end{equation}

Now, $\phi_{\rm other}$ can be re-written as follows:
\begin{equation}
	\phi_{\rm other} = \frac{1}{p} \sum_{k=1}^p  \frac{1}{L-1} \sum_{\ell' \neq l} a^{\ell'}_{t-k}.
\end{equation} 
Then, by the law of large numbers we can express this in terms of the sample average:
\begin{eqnarray}
\begin{aligned}
	\phi_{\rm other} =& \frac{1}{p} \sum_{k=1}^{p} \left< a^{\ell'}_{t-k} \right>_{\ell'}, \\
	=&  \frac{1}{p} \sum_{k=1}^{p} \bar{a},\\
	=&\bar{a}.
\end{aligned}
\end{eqnarray}
Hence $\phi_{\rm other} \to \bar{a}$ as $L \to \infty$. 
Indeed, we can further see that $\bar{a} = y$.
Since there are no terms containing links other than $\ell$ in $\phi_{\rm self}$, then we can conclude that the conditional probability is such that, in the same limit $L \to \infty$,
\begin{equation}
\text{Prob}(a^{\ell}_t = 1 | \{ \{\underline{\underline{A}}_{s}\}_{s = t}^{t-p})  \to \text{Prob}(a^{\ell}_t = 1 | \{a^{\ell}_{s}\}_{s = t}^{t-p}),
\end{equation}
and so any memory of other links is lost.
To show that this is equivalent to a DARN($p$) network
we need only look at the conditional probability of obtaining a link in
such a network with memory strength $\bar{q}$, memory length $p$,
link density $\bar{y}$ and adjacency matrix $\underline{\underline{E}}_t$ with observed values $\{e^{\ell}_{t}\}$:
\begin{equation}
	\text{Prob}(e^{\ell}_t = 1 | \{ \{\underline{\underline{E}}_{s}\}_{s = t}^{t-p}) = (1- \bar{q}) \bar{y} \, + \frac{\bar{q}}{p} \sum_{k=1}^{p} e^{\ell}_{t-k}.
\end{equation}
Now, by setting the values of $\bar{q}$ and $\bar{y}$, in terms of the values $q$, $y$ and $c$ from 
the CDARN($p$) model, to be 
\begin{eqnarray}
\begin{aligned}
	\bar{q} =& q(1-c), \\
	\bar{y} =& y,
\end{aligned}
\end{eqnarray}
we obtain that 
\begin{equation}
	\text{Prob}(e^{\ell}_t = 1 | \{ \{\underline{\underline{A}}_{s}\}_{s = t}^{t-p}) = \text{Prob}(a^{\ell}_t = 1 | \{ \{\underline{\underline{A}}_{s}\}_{s = t}^{t-p}).
\end{equation}
Hence the UCC model is precisely a DARN($p$) model in the limit of 
$L \to \infty$.


\subsection{\cg{MLE of the CDARN(p) model}}
\label{sm_mle}

\cg{The CDARN(p) model can be estimated by the maximum likelihood method. 
First of all, consider the vectorization $\underline{\underline{X}}_t\equiv\{a^\ell_t\}^{\ell=1,...,L}$, with $L$ 
the number of links on the backbone, of the adjacency matrix 
$\{a^{ij}_t\}^{(i,j)\in B}$ of the network snapshot at time $t$. That is, 
$\{\underline{\underline{X}}_t\}_{t=1,...,T}$ describes the binary random sequences associated 
with the dynamics of the $L$ links on the backbone. Then, the log-likelihood 
of data (by conditioning on the first $p$ observations) \piero{under CDARN(p)}, as defined in Section 
\ref{empcdarn}, reads as:
\begin{equation}\label{llcdarnx}
\begin{split}
\mathbb{L}&(q,c,y) \equiv \log \mathbb{P}(\{\underline{\underline{X}}_t\}_{t=p+1,...,T}\vert \{\underline{\underline{X}}_s\}_{s=1,...,p},q,c,y) \\
& = \sum_{t,l}\log \left[ q((1-c)D_t^\ell+c C_t^\ell)+(1-q)y^{X_t^\ell}(1-y)^{1-X_t^\ell} \right],
\end{split}
\end{equation}
where $t$ runs from $p+1$ to $T$, while $\ell$ from $1$ to $L$, with
$$
D_t^\ell = \sum_{\tau=1}^p z_\tau \delta(X_t^\ell,X_{t-\tau}^\ell),
$$
$$
C_t^\ell =\sum_{\ell'\neq \ell}\lambda^{\ell\ell'} \sum_{\tau=1}^p z_\tau \delta(X_t^\ell,X_{t-\tau}^{\ell'}),
$$
where $\delta(a,b)$ is the Kronecker delta, taking value equal to one
if $a=b$, zero otherwise, $z_\tau$ is the probability of picking
$\tau$ in the range of integers $(1,...,p)$ (it is $z_\tau=1/p$ if we
assume uniform probability), and $\lambda^{\ell\ell'}$ is:
\begin{enumerate}
\item $\lambda^{\ell\ell'}=0$ $\forall \ell,\ell'=1,....,L$ with $\ell\neq \ell'$, for the {\it no cross correlation (NCC)} coupling model;
\item $\lambda^{\ell\ell'}=1/\vert\partial_B \ell \vert$ with $\vert\partial_B \ell \vert$ the number of neighbours of link $\ell$ if $\ell'\in\partial_B \ell$, zero otherwise, for the {\it local cross correlation (LCC)} coupling model;
\item $\lambda^{\ell\ell'}=1/(L-1)$ with $\ell\neq \ell'$, for the {\it uniform cross correlation (UCC)} coupling model.
\end{enumerate}
}

\cg{The MLE of the CDARN(p) model can be then obtained by maximising the
log-likelihood in Eq.~(\ref{llcdarnx}), or, equivalently, by solving the
following system of non-linear equations:
\begin{equation}\label{mle_cdarnp_eq}
\begin{cases}
\frac{\partial \mathbb{L}}{\partial y}&= \sum_{t,\ell} \frac{2X_t^\ell-1}{q((1-c)D_t^\ell+c C_t^\ell)+(1-q)y^{X_t^\ell}(1-y)^{1-X_t^\ell}}=0,\\
\frac{\partial \mathbb{L}}{\partial q}&= \sum_{t,\ell} \frac{((1-c)D_t^\ell+c C_t^\ell)- y^{X_t^\ell}(1-y)^{1-X_t^\ell}}{q((1-c)D_t^\ell+c C_t^\ell)+(1-q)y^{X_t^\ell}(1-y)^{1-X_t^\ell}}=0,\\
\frac{\partial \mathbb{L}}{\partial c}&= \sum_{t,\ell} \frac{C_t^\ell-D_t^\ell}{q((1-c)D_t^\ell+c C_t^\ell)+(1-q)y^{X_t^\ell}(1-y)^{1-X_t^\ell}}=0.
\end{cases}
\end{equation}
}
\piero{The system of non-linear equations can be solved iteratively by adopting an {\it iterative proportional fitting procedure}. This consists in solving one by one each equation for each parameter, but conditioning on the values of the other parameters, up to convergence. Such procedure can be initialized randomly in the parameter space, however a natural initialization for the parameter $y$ is the average link density of the network. For further details on the method see also \cite{mazzarisi2019dynamic}. }

\subsection{\piero{MLE of the {\it heterogeneous} CDARN(p) model}}
\label{sm_hetero_mle}
\piero{In the case of link-specific parameters $y^\ell$ or $q^\ell$ with $\ell=1,\ldots,L$, the log-likelihood of data under CDARN(p) with heterogenous parameters is generalized quite naturally as
\begin{equation}\label{llcdarnxHy}
\begin{split}
\mathbb{L}&(q,c,\underline{y}) \equiv \log \mathbb{P}(\{\underline{\underline{X}}_t\}_{t=p+1,...,T}\vert \{\underline{\underline{X}}_s\}_{s=1,...,p},q,c,\underline{y}) \\
& = \sum_{t,\ell}\log \left[ q((1-c)D_t^\ell+c C_t^\ell)+(1-q)(y^\ell)^{X_t^\ell}(1-y^\ell)^{1-X_t^\ell} \right],
\end{split}
\end{equation}
and
\begin{equation}\label{llcdarnxHq}
\begin{split}
\mathbb{L}&(\underline{q},c,y) \equiv \log \mathbb{P}(\{\underline{\underline{X}}_t\}_{t=p+1,...,T}\vert \{\underline{\underline{X}}_s\}_{s=1,...,p},\underline{q},c,y) \\
& = \sum_{t,\ell}\log \left[ q^\ell((1-c)D_t^\ell+c C_t^\ell)+(1-q^\ell)y^{X_t^\ell}(1-y)^{1-X_t^\ell} \right],
\end{split}
\end{equation}
respectively for $y^\ell$ and $q^\ell$. Then, similarly to before, the MLE of the CDARN(p) model with heterogenous parameters is obtained by solving
\begin{equation}\label{mle_cdarnp_eq_Hy}
\begin{cases}
\frac{\partial \mathbb{L}}{\partial y^\ell}&= \sum_{t} \frac{2X_t^\ell-1}{q((1-c)D_t^\ell+c C_t^\ell)+(1-q)(y^\ell)^{X_t^\ell}(1-y^\ell)^{1-X_t^\ell}}=0,\\
\frac{\partial \mathbb{L}}{\partial q}&= \sum_{t,\ell} \frac{((1-c)D_t^\ell+c C_t^\ell)- (y^\ell)^{X_t^\ell}(1-y^\ell)^{1-X_t^\ell}}{q((1-c)D_t^\ell+c C_t^\ell)+(1-q)(y^\ell)^{X_t^\ell}(1-y^\ell)^{1-X_t^\ell}}=0,\\
\frac{\partial \mathbb{L}}{\partial c}&= \sum_{t,\ell} \frac{C_t^\ell-D_t^\ell}{q((1-c)D_t^\ell+c C_t^\ell)+(1-q)(y^\ell)^{X_t^\ell}(1-y^\ell)^{1-X_t^\ell}}=0,
\end{cases}
\end{equation}
and
\begin{equation}\label{mle_cdarnp_eq_Hq}
\begin{cases}
\frac{\partial \mathbb{L}}{\partial y}&= \sum_{t,\ell} \frac{2X_t^\ell-1}{q^\ell((1-c)D_t^\ell+c C_t^\ell)+(1-q^\ell)y^{X_t^\ell}(1-y)^{1-X_t^\ell}}=0,\\
\frac{\partial \mathbb{L}}{\partial q^\ell}&= \sum_{t} \frac{((1-c)D_t^\ell+c C_t^\ell)- y^{X_t^\ell}(1-y)^{1-X_t^\ell}}{q^\ell((1-c)D_t^\ell+c C_t^\ell)+(1-q^\ell)y^{X_t^\ell}(1-y)^{1-X_t^\ell}}=0,\\
\frac{\partial \mathbb{L}}{\partial c}&= \sum_{t,\ell} \frac{C_t^\ell-D_t^\ell}{q^\ell((1-c)D_t^\ell+c C_t^\ell)+(1-q^\ell)y^{X_t^\ell}(1-y)^{1-X_t^\ell}}=0,
\end{cases}
\end{equation}
respectively.}

\subsection{\cg{Link prediction in the CDARN(p) model}}
\label{sm_forecast}

\cg{Once estimated on data by solving the MLE problem 
(\ref{mlecdarn}), the CDARN(p) model can be used 
for link prediction: assume that we observe a temporal network up to time 
$t$ and are asking for the prediction of the network snapshot at time 
$t+1$ (by using only the information up to time $t$). The one-step-ahead 
{\it forecast} (or {\it prediction}) of link $\ell$ is defined as
\begin{equation}\label{lp_cdarn}
\begin{split}
S_{t+1}^\ell & \equiv \mathbb{P}(X_{t+1}^\ell=1\vert \{\underline{\underline{X}}_s\}_{s=t,t-1,...,t-p+1},q,c,y)=\\ & = q((1-c)\tilde{D}_{t+1}^\ell+c \tilde{C}_{t+1}^\ell)+(1-q)y,
\end{split}
\end{equation}
with
$$
\tilde{D}_{t+1}^ell = \sum_{\tau=1}^p z_\tau \delta(1,X_{t-\tau+1}^\ell),
$$
$$
\tilde{C}_{t+1}^\ell =\sum_{\ell'\neq \ell}\lambda^{\ell\ell'} \sum_{\tau=1}^p z_\tau \delta(1,X_{t-\tau+1}^{\ell'}).
$$
The one-step ahead forecast in Eq.~(\ref{lp_cdarn}) is described by a
real value in the unit interval, representing the probability
projected at time $t+1$ of observing a link, then the prediction
itself, namely the binary value $\tilde{X}_{t+1}^\ell\in \{0,1\}$, is
obtained according to some threshold value. The time series of
forecasts $\{S_t^\ell\}$, together with the realisations $\{X_t^\ell\}$,
allow us to characterise the forecasting performance of the model by
using some binary classifier. A possibility is constructing the
Receiving Operating Characteristic (ROC) curve
\cite{hastie2009elements}, which is the plot of the True
Positive Rate (TPR) ({\it sensitivity}) against the False Positive
Rate (FPR) ({\it specificity}) at various threshold values of the
link probability. In particular, the threshold values are selected
implicitly by the inputs themselves: by moving from zero to one in the
unit interval, each time the sensitivity is increasing or the
specificity is decreasing, the corresponding value is considered as a
threshold. In practical terms, the better the model performs in the
forecasting, the higher the associated ROC curve is in the unit
square, or, equivalently, the larger the Area Under the Curve (AUC).
}

\subsection{\piero{Network statistics}}
\label{cdarn_validation}
\begin{figure*}
\centering
\includegraphics[width=0.99\textwidth]{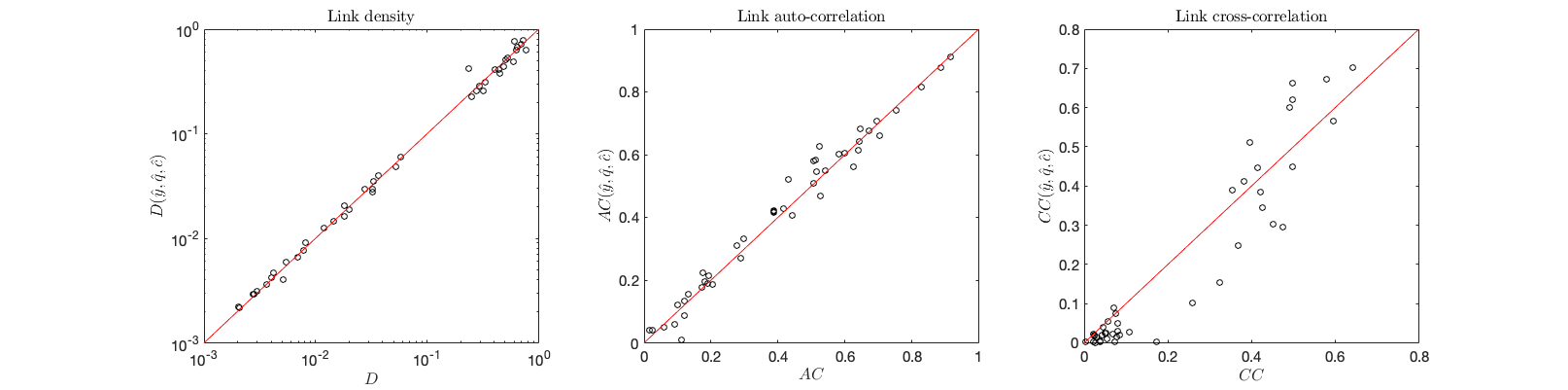}
\caption{\piero{Scatter plots of the mean link density (left), auto-correlation (middle), and cross-correlation (right) (at lag equal to one) of the empirical temporal networks built with the datasets described in the main text vs. the same statistics computed on simulations of the CDARN(1) model (LCC) with constant parameters, which are estimated on the temporal networks themselves.}}
\label{figStats}
\end{figure*}

\piero{Once estimated on data, the CDARN(p) model in the standard version with constant parameters and for a specific coupling specification, captures the average link density of a temporal network, together with the average auto- and cross-correlations of links, in particular the cross-correlations of links interacting on the backbone, \eg all links that are incident to the same nodes (neighbours links) for the LCC specification. Thus, the observed statistics match (on average) their expectations according to the CDARN(p) model, eventually computed by using simulations. For validation purpose, this is shown in Fig.~\ref{figStats} for all the datasets described in Section \ref{empcdarn} by considering the CDARN(1)-LCC model with constant parameters, which are estimated on data by using maximum likelihood methods described above.}

\piero{More interestingly, other networks statistics that are not explicitly described by CDARN models, can be computed within the proposed framework, in particular the inter-event time between the occurrence of two subsequent links between two nodes at two different times. For instance, for temporal networks of human communication, it has been observed that the duration between two contacts is often bursty, deviating from a uniform distribution expected by some memoryless process \cite{Holme_rev12}. This quantity is of great interest, \eg, when studying any spreading process taking place on temporal networks, such as diffusion as in the present work. In our setting, the inter-event time can be defined as the number $\tau$ of observed time snapshots $X_{t+1}^\ell,\ldots,X_{t+\tau}^\ell$ for which the generic link $\ell$ is zero, after an observation $X_t^\ell=1$. The distribution of $\tau$ conditional to the observation $X_t^\ell=1$ for the CDARN(1) model (with constant parameters) is equivalent to compute the following joint probability
\be\label{ptauCDARN}
\begin{aligned}
p\left(\tau\vert X_t^\ell=1\right)\equiv P(X_{t+1}^\ell=0,\ldots,X_{t+\tau}^\ell=0\vert X_t^\ell=1) =\\
= P(X_{t+1}^\ell=0\vert X_t^\ell=1)\prod_{i=1}^{\tau-1}P(X_{t+1+i}^\ell=0\vert X_{t+i}^\ell=0)=\\
=\left(qc\sum_{\ell'\neq \ell}\lambda^{\ell\ell'}\delta(0,X_{t}^{\ell'})+\beta\right)\times\\\times\prod_{i=1}^{\tau-1}\left[q\left((1-c)+c\sum_{\ell'\neq \ell}\lambda^{\ell\ell'}\delta\left(0,X_{t+i}^{\ell'}\right)\right)+\beta\right],
\end{aligned}
\ee
with $\beta=(1-q)(1-y)$.} 

\bigskip
\piero{In general, a closed form solution cannot be obtained because of non-diagonal interaction terms mediated by the couplings matrix $\lambda$. Except for the NCC specification of the model, \ie the DARN(1) model, which sets to zero the cross-correlations, \ie $c=0$. In this case, the probability distribution of the inter-event time is
\be\label{ptauDARN}
p(\tau)= e^{-\frac{\tau-1}{\sigma}},\:\tau=1,2,\ldots,
\ee
with $\sigma = \left(\log\frac{1}{q+\beta}\right)^{-1}$, as follows from simple computations.}

\begin{figure*}
\centering
\includegraphics[width=0.49\textwidth]{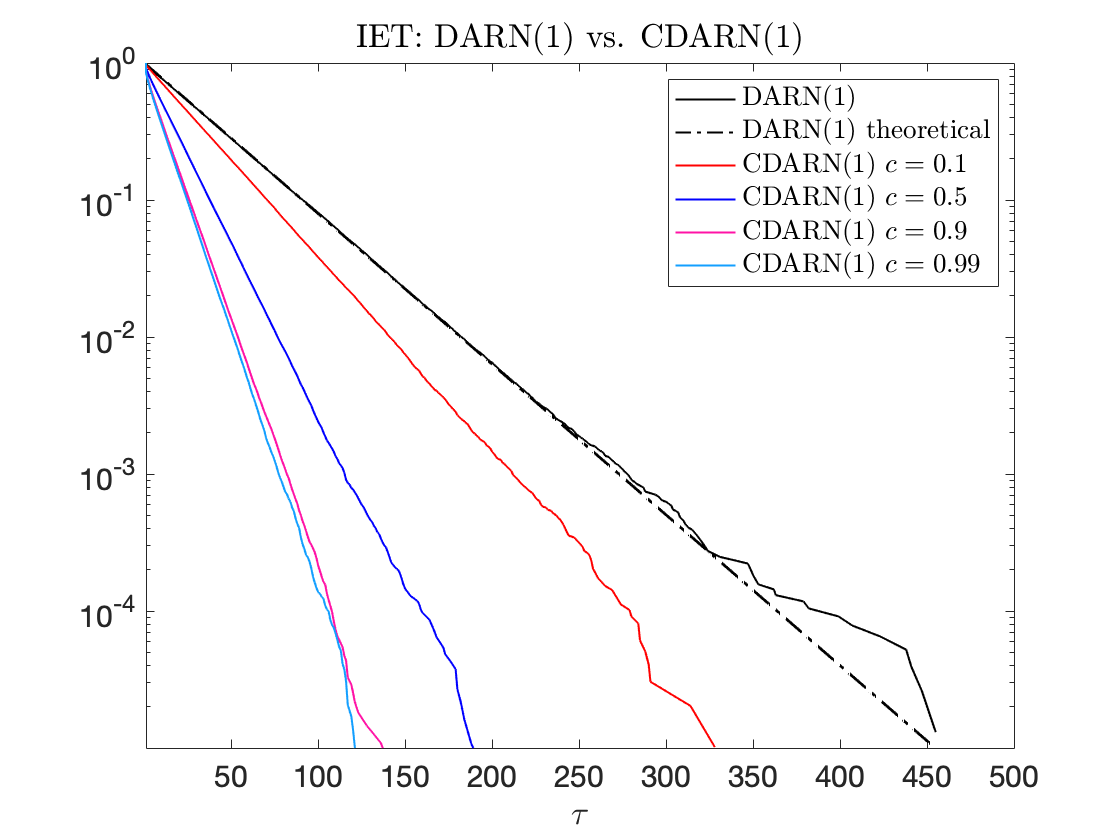}
\includegraphics[width=0.49\textwidth]{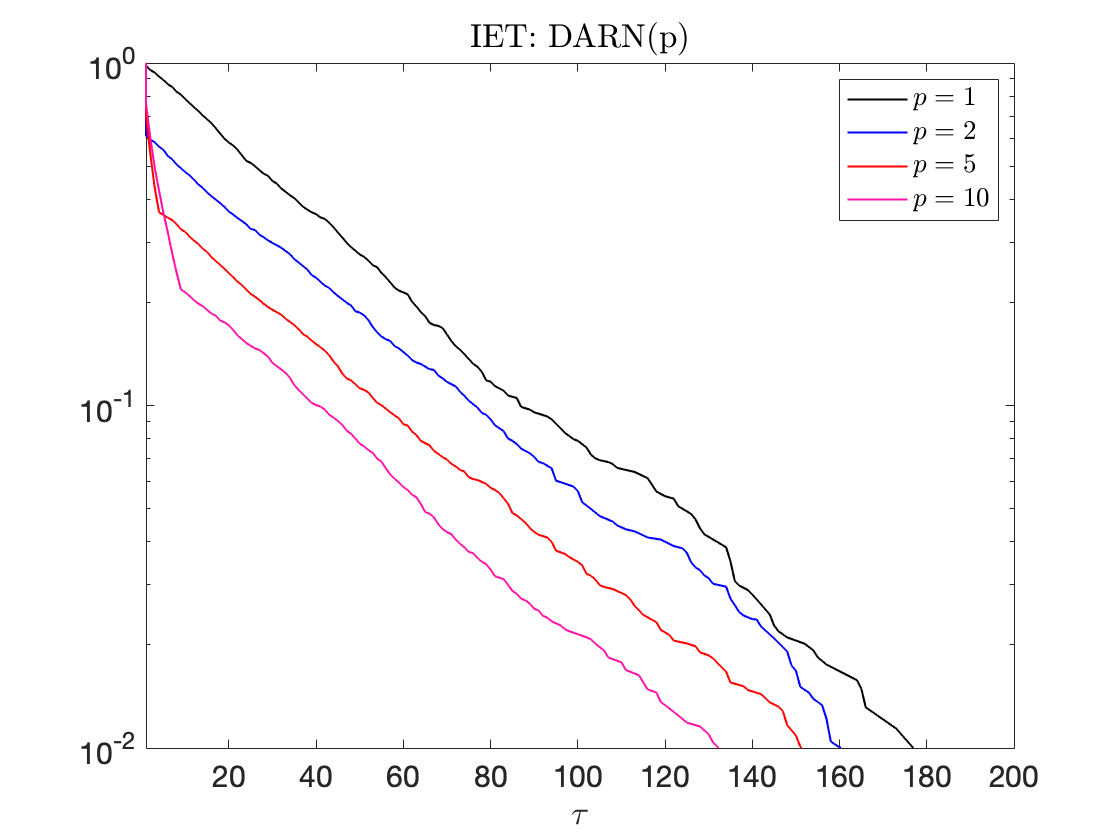}
\caption{\piero{Comparison of the Inter-Event Time (IET) distributions between DARN(1) and CDARN(1) (LCC) models (left), and of the DARN(p) models for different $p$ (right). IET is obtained by using simulations of time series of length $T=10^5$ of the models, with $y=0.1$, $q=0.75$, and $c$ as indicated in the plots (for CDARN).}}
\label{figIETsim}
\end{figure*}

\piero{Notice that the probability distribution (\ref{ptauDARN}) of the DARN(1) model represents an upper bound for the CDARN(1) model: when $c>0$ in Eq. (\ref{ptauCDARN}), there exists always a probability larger than zero of copying one past neighbour link (whatever the coupling matrix $\lambda$), instead of copying the past itself, \ie a zero, with probability one. This reduces the probability of observing a number $\tau$ of successive zeros, thus resulting in an ({\it approximate}) exponential distribution of inter-event times with a time scale smaller than $\sigma$, see the left panel of Fig.~\ref{figIETsim}. Finally, the non-Markovian case $p>1$ is not analytically tractable as long as $p$ increases further and further. However, for the NCC specification of the CDARN(p) model, \ie DARN(p), it is easy to gather that $p(\tau)= O\left[ \exp(-(\tau-p)/\sigma)\right]$ when $\tau\gg p$, by following similar computations leading to Eq. (\ref{ptauCDARN}). This is confirmed numerically in the right panel of Fig.~\ref{figIETsim}. We can conclude that the inter-event time distribution of the CDARN(p) model is {\it approximate} exponential with a time scale equal or smaller than $\sigma$.}

\begin{figure*}
\centering
\includegraphics[width=0.99\textwidth]{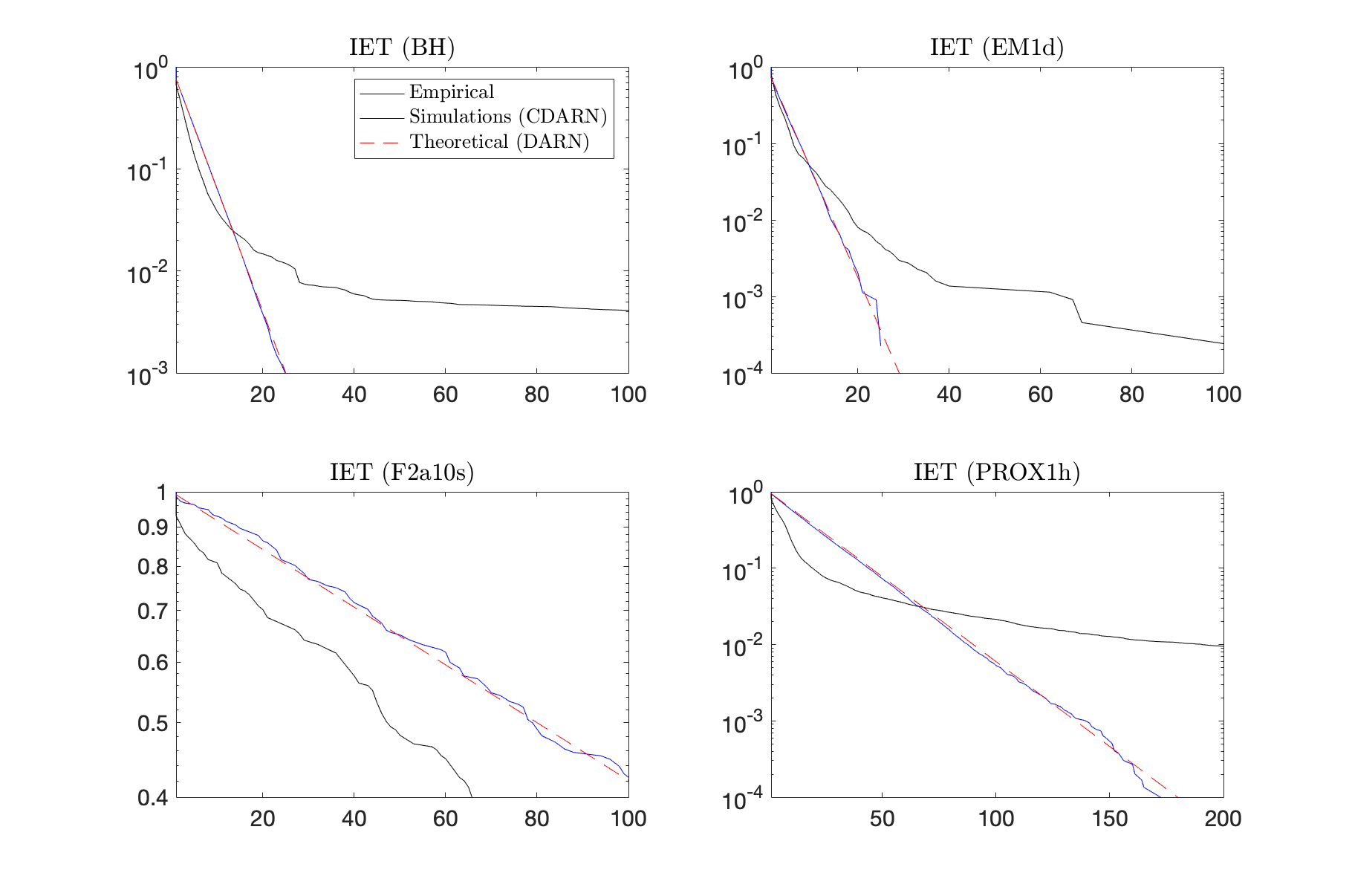}
\caption{\piero{Inter-Event Time (IET) distribution for four temporal networks, \ie BH, EM1d, F2a10s, and PROX60min, compared with IET distributions for the CDARN(1) model with LCC specification based on numerical simulations, and with the theoretical IET distribution for the DARN(1) model, which can be computed analytically as explained in the main text.}}
\label{figIETemp}
\end{figure*}

\piero{Finally, in Fig.~\ref{figIETemp} we show the empirical distribution of the inter-event time of four real-world temporal networks, compared with the corresponding distribution for the CDARN(1) model with $LCC$ coupling specification, obtained numerically by means of simulations, and with the theoretical one (\ref{ptauDARN}) for the DARN(1) model. In both cases, the parameters of the model has been obtained by maximum likelihood estimation. In all cases, the CDARN(1) can be seen as an approximation of the empirical distributions of the inter-event time for small $\tau$, while the {\it fatter} tails (excluding the football network) are not captured by the model.} 

\vfill



%

\end{document}